\documentclass{IEEEcsmag}

\usepackage[colorlinks,urlcolor=blue,linkcolor=blue,citecolor=blue]{hyperref}
\expandafter\def\expandafter\UrlBreaks\expandafter{\UrlBreaks\do\/\do\*\do\-\do\~\do\'\do\"\do\-}
\usepackage{upmath,color}
\usepackage[table,xcdraw]{xcolor}
\usepackage{pdflscape}
\usepackage{float}
\usepackage{graphicx}

\jvol{XX}
\jnum{XX}
\paper{8}
\jmonth{October}
\jname{Publication Name}
\jtitle{HIKMA: Human-Inspired Knowledge by Machine Agents}
\pubyear{2025}

\begin{document}


\sptitle{Feature Article: Human-Inspired Knowledge by Machine Agents}

\title{HIKMA: Human-Inspired Knowledge by Machine Agents through a Multi-Agent Framework for Semi-Autonomous Scientific Conferences}

\author{Zain Ul Abideen Tariq}
\affil{College of Science and Engineering, Hamad Bin Khalifa University, Qatar}

\author{Mahmood Al-Zubaidi}
\affil{College of Science and Engineering, Hamad Bin Khalifa University, Qatar}

\author{Uzair Shah}
\affil{College of Science and Engineering, Hamad Bin Khalifa University, Qatar}

\author{Marco Agus}
\affil{College of Science and Engineering, Hamad Bin Khalifa University, Qatar}

 \author{Mowafa Househ}
\affil{College of Science and Engineering, Hamad Bin Khalifa University, Qatar}


\markboth{FEATURE ARTICLE}{FEATURE ARTICLE}

\begin{abstract}\looseness=-1 HIKMA Semi-Autonomous Conference is the first experiment in reimagining scholarly communication through an end-to-end integration of artificial intelligence into the academic publishing and presentation pipeline. This paper presents the design, implementation, and evaluation of the HIKMA framework, which includes AI-dataset curation, AI-based manuscript generation, AI-assisted peer review, AI-driven revision, AI conference presentation, and AI archival dissemination. By combining language models, structured research workflows, and domain safeguards, HIKMA shows how AI can support—not replace—traditional scholarly practices while maintaining intellectual property protection, transparency, and integrity. The conference functions as a testbed and proof of concept, providing insights into the opportunities and challenges of AI-enabled scholarship. It also examines questions about AI authorship, accountability, and the role of human-AI collaboration in research.
\medskip
\noindent\textbf{Corresponding author: Mowafa Househ} mhouseh@hbku.edu.qa
\end{abstract}

\maketitle

\section{Introduction}
\label{sec:intro}
\vspace{0.2cm}

The pursuit of artificial intelligence has long been guided by the question of whether machines can reach the level of human intelligence in domains that demand reasoning, creativity, and judgment. Academic research and scholarly writing represent one of the most rigorous tests of this capability: they require not only the generation of coherent text, but also the integration of evidence, critical analysis, peer review, and ethical responsibility \cite{floridi2020gpt, gil2022will, salvagno2023can}. Recent advances in large language models and generative systems demonstrate that AI can already draft manuscripts, simulate peer review, and generate presentations \cite{auernhammer2020human, afzal2025chatgpt}. Yet the transition from promising prototypes to trustworthy scholarly infrastructure remains incomplete.

Modern AI can produce text that resembles academic discourse, but the processes surrounding it lack mechanisms to ensure trustworthiness, safeguard intellectual property, and provide audit-ready transparency \cite{donovan2018algorithmic, li2023trustworthy}. Without verifiable provenance and accountability, AI-driven scholarly workflows risk undermining the very integrity of academic communication they aim to support. The challenge is therefore not only technical, but also institutional and ethical: how to design systems that allow automation while remaining consistent with the principles of reliability, rights protection, and independent verification \cite{weidinger2021ethical}.

This paper seeks to address this gap by presenting an actionable blueprint for the first-ever semi-autonomous scholarly communication. The blueprint is designed to enable researchers, institutions, and publishers to integrate AI into the research lifecycle in a manner that preserves intellectual property, maintains transparency, and supports independent audits \cite{yoo2024evolving}. By formalizing the structure and traceability of AI contributions, the approach transitions from ad hoc experimentation to a sustainable and accountable model of machine-assisted scholarship.

The contributions of this work are threefold. First, we propose an end-to-end architecture that implements human-supervised, AI-driven semi-autonomous scholarly communication. This pipeline spans the full research lifecycle, from AI-powered dataset search and selection, through AI-assisted draft paper generation using large language models, AI peer review conducted by independent agent reviewers, iterative AI-led revision cycles, and final camera-ready acceptance and archiving. Second, we introduce an AI-traceable audit trail that captures each stage of the process—including dataset choices, manuscript versions, AI-generated reviewer comments, revision histories, and presentation materials—ensuring transparency and enabling independent verification. Third, we present evaluation metrics and results from the Human-Inspired Knowledge by Machine Agents (HIKMA) experiment, where this AI-enabled pipeline was demonstrated at scale: sixty draft papers were generated by AI, reviewed by AI agents, and revised through automated feedback integration. Thirty manuscripts progressed to final acceptance, AI-generated presentation decks, avatar-based narration, video archiving, and public dissemination. Together, these contributions outline a pathway for AI to move beyond narrow task automation and toward functioning as a reliable, auditable partner in the production, review, and dissemination of scientific knowledge.

\subsection{Scope and claims}

This work is scoped around the end-to-end workflow of an AI-driven academic conference, with a focus on demonstrating how the \textit{AI Scholar Frontier} tool can support autonomous scholarly communication. \textit{AI Scholar Frontier} is designed as an integrated research assistant that combines large language models, structured templates, and workflow orchestration to generate, review, revise, and present academic content. Rather than addressing isolated tasks, the system implements the entire scholarly communication pipeline, enabling automation with transparency and auditability. The workflow encompasses the following stages:

\begin{enumerate}
    \item \textbf{Dataset Intake.} \textit{AI Scholar Frontier} initiates the process by searching repositories such as Kaggle and other open platforms. A total of 60 datasets were searched and screened for sufficiency and diversity before being selected as inputs for paper generation.

    \item \textbf{Paper Generation.} Using its manuscript generation module, \textit{AI Scholar Frontier} produced 60 draft papers. Each draft followed a conventional academic structure, including an abstract, introduction, methodology, results, discussion, and references.

    \item \textbf{Peer Review.} The peer review module enabled two independent AI agents to evaluate each draft, resulting in 120 reviews. Reviews considered originality, methodological rigor, clarity, reproducibility, significance, and ethical integrity.

    \item \textbf{Revision and Response.} The revision process was conducted in two distinct AI-assisted cycles. In the first cycle, \textit{AI Scholar Frontier} analyzed reviewer feedback and produced 15 revised manuscript drafts that directly incorporated the suggested changes. In the second cycle, the system generated 30 formal response letters—each corresponding to a reviewed manuscript—explicitly documenting how reviewer comments were addressed, clarified, or rebutted. This dual-phase revision ensured both content improvement and transparent communication, aligning with academic standards for peer-reviewed publication.

    \item \textbf{Camera-Ready Acceptance and Archiving.} Accepted manuscripts were finalized, version-controlled, and archived with metadata and timestamps. A total of 30 manuscripts advanced to this stage.

    \item \textbf{Slide Synthesis.} \textit{AI Scholar Frontier} converted accepted manuscripts into structured presentation slides, with figures, charts, and highlights automatically generated. This produced 30 presentation decks.

    \item \textbf{Avatar Presentation.} Narration scripts were generated using \textit{AI Scholar Frontier}, and avatar-based presentations were rendered using the HeyGen platform \cite{heygen2022}. This enabled fully automated delivery of scholarly talks, substituting human presenters with AI-generated avatars that maintained professional tone and visual consistency.

    \item \textbf{Archival and Publishing.} Final video presentations—produced via HeyGen \cite{heygen2022}—were recorded and archived alongside all associated materials, including AI-generated papers, slides, and response letters. These assets were published on the conference website for open access, ensuring transparency, reproducibility, and broad dissemination of the AI-assisted scholarly outputs.
\end{enumerate}

\begin{figure*}[h]
  \centering
  \includegraphics[width=\linewidth]{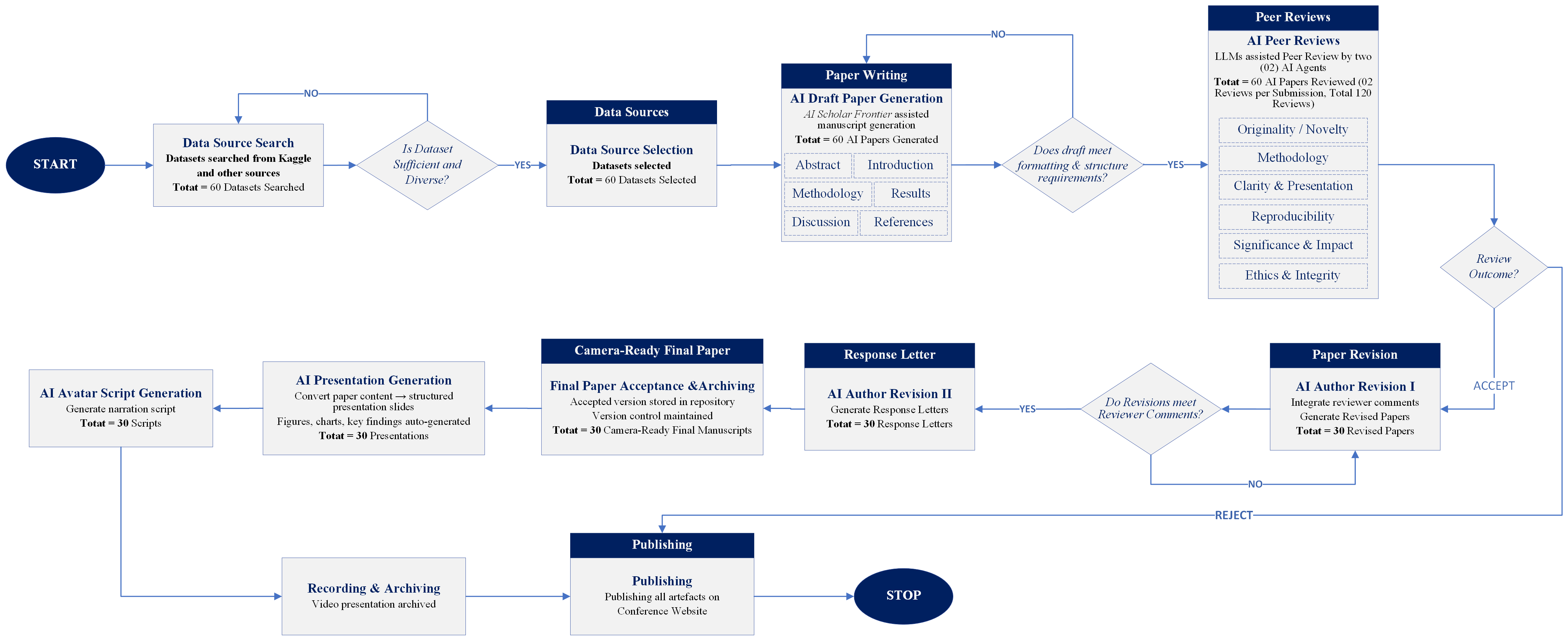}
  \caption{HIKMA process pipeline from data intake to avatar presentations.}
  \label{fig:pipeline}
\end{figure*}

The scope is therefore defined by the scholarly communication lifecycle as executed through \textit{\textit{AI Scholar Frontier}}: dataset intake $\rightarrow$ paper generation $\rightarrow$ peer review $\rightarrow$ revision/response $\rightarrow$ camera-ready $\rightarrow$ slide synthesis $\rightarrow$ avatar presentation $\rightarrow$ archival as shown in Figure \ref{fig:pipeline}. This pipeline, validated through the HIKMA experiment and demonstrated on the website, serves as a reference model for evaluating how AI can function as an autonomous partner in research, writing, and dissemination.

\section{Related Works}
\vspace{0.2cm}

The automation of scientific research workflows has become an increasingly active area of study, with advances in natural language processing (NLP) and generative AI enabling tools that support hypothesis analysis, research writing, and academic communication. This section reviews prior work and systems relevant to the HIKMA experiment, emphasizing their limitations and the way in which HIKMA addresses these shortcomings.

\subsection*{AI-Assisted Hypothesis Analysis and Generation}

Early attempts to computationally model scientific discovery relied on symbolic reasoning and structured knowledge bases. Swanson’s concept of \textit{undiscovered public knowledge} \cite{swanson1986undiscovered} and the \textit{Robot Scientist} project \cite{sparkes2010towards} pioneered automated hypothesis generation, but they were limited by constrained domains and the need for manually curated inputs. More recent advances leverage large language models (LLMs) to scale hypothesis generation. McCall \cite{mccall2025ai} explored how AI can automate hypothesis generation, while Salvagno et al. asked whether AI can write full scientific papers \cite{salvagno2023can}. The same question was explored by Nathani et al. by carrying out the comparative analysis of human-written and AI-generated scientific writings \cite{nathani2025can}. However, these works typically address hypothesis suggestion in isolation, without extending into writing, review, or dissemination. The HIKMA experiment expands beyond this narrow scope by connecting hypothesis intake directly to a reproducible end-to-end scholarly workflow.

\subsection*{AI for Research Writing and Drafting}

Generative LLMs such as GPT-3 and GPT-4 have shown strong capabilities for producing structured academic text \cite{floridi2020gpt, afzal2025chatgpt}. Tools like \textit{Elicit} \cite{elicit2023, whitfield2023elicit}  assist with literature exploration and argument extraction, while platforms such as \textit{Writefull} \cite{writefull2022, bute2025writefull} and \textit{Scholarcy} \cite{scholarcy2022, bui2024decoding} focus on language refinement and summarization. Yet these systems largely operate at the drafting or polishing stage, without integrating peer review or lifecycle management. Their scope remains tool-specific, with little capacity for traceability or auditability. Conversely, HIKMA integrates drafting, revision, and archiving, thereby transforming isolated support functions into a coherent, auditable pipeline.

\subsection*{AI Scientists and Co-Researchers}

Recent developments in the field of autonomous research systems have aimed to position artificial intelligence as a co-investigator capable of contributing to scientific discovery. Google’s \textit{Co-Scientist} tool \cite{co_scientist2024} has demonstrated capabilities in automating hypothesis testing and experiment design, enabling researchers to iteratively query datasets and models. Similarly, \textit{Sakana AI Scientist} \cite{sakana2023} focuses on end-to-end autonomous research agents capable of designing and testing scientific hypotheses in computational environments and producing a pre-print of the paper. Although both represent significant steps toward autonomous scientific discovery, their emphasis remains on hypothesis exploration and experimental reasoning. They do not incorporate the full scholarly communication process, particularly peer review, response to feedback, and dissemination through presentations. 

HIKMA platform addresses this gap by placing hypothesis exploration within a complete research-to-publication pipeline, ensuring that discoveries are communicated, reviewed, and archived in a reproducible manner.

A related initiative is the \textit{Agents for Science} open conference developed at Stanford University~\cite{stanford2025agents4science}, which explores the role of autonomous AI agents as both authors and reviewers of scientific papers. The project demonstrates a live experimental environment where large language model-based agents generate research manuscripts, conduct automated peer review, and provide feedback loops under human oversight. The Stanford framework serves as an open testbed for studying agent cooperation, role specialization, and evaluation of multi-agent research systems. It represents a key development for understanding how AI can participate directly in the academic peer review process and contribute to the production of scientific literature.

However, despite its innovation, the \textit{Agents for Science} framework primarily focuses on the interaction between authoring and reviewing agents within a limited submission–review–decision cycle. It does not extend into the post-review lifecycle of scholarly communication, such as revision management, response-to-review generation, or camera-ready preparation. Nor does it include dissemination mechanisms like slide synthesis, avatar-based presentation, or archival integrity verification. In addition, governance mechanisms—such as provenance tracking, authorship transparency, dataset licensing, and auditability-are not deeply embedded in its workflow. 

The HIKMA experiment builds upon and extends these foundations by implementing a complete, auditable research-to-publication pipeline. In HIKMA, AI agents do not only author and review manuscripts but also handle structured revisions, produce camera-ready versions, generate presentation slides, create narrated avatar-based videos, and archive outputs with version-controlled provenance metadata. Through this integrated lifecycle, HIKMA transforms the concept of AI-assisted conferences into a reproducible, transparent scholarly platform that embeds accountability, ethical compliance, and end-to-end communication within the automation process.

Another recent work by Zhang et. al \cite{zhang2025aixiv} presents one of the most advanced implementations of this idea. It describes an AI-driven environment that automates the synthesis of the literature, the integration of data sets, the generation of hypothesis, the generation of articles, and the automated review through interconnected research agents. These AI scientists are capable of identifying gaps in existing literature, designing basic experimental frameworks, and publishing findings in open-access repositories. The proposed model emphasizes open collaboration between AI systems and human researchers, with a focus on accessibility, dataset sharing, and open knowledge exchange.  

Although this system represents a step significant advancement toward scalable AI-assisted discovery, it remains primarily centered on hypothesis generation and dataset-level reasoning. It does not incorporate the procedural and communicative stages that define a complete scholarly lifecycle. Specifically, it does not include structured revision and response loops, or post-acceptance dissemination such as slide synthesis, avatar-based presentation, and archival verification. Furthermore, the governance aspects—such as version control, ethical audit trails, authorship attribution, and public traceability through provenance graphs—are not fully integrated within its framework. 

The HIKMA 2025 workflow builds on prior systems by implementing a full research-to-publication process with version tracking and verification at each stage, ensuring that outputs are produced, reviewed, and archived in traceable form. It connects manuscript drafting and peer reviews to revision, camera-ready preparation, and dissemination in a continuous, auditable process. Through this process, it transforms AI-driven discovery from an experimental automation framework into a reproducible scholarly communication system that produces knowledge and validates it within the same workflow, communicates, and archives it within established academic standards.

\subsection*{AI for Peer Review and Quality Assessment}

Studies have examined the feasibility of AI-assisted peer review. Efforts such as Checco et al. \cite{checco2021ai} and Donker et al. \cite{donker2023dangers} explored how large language models could provide structured feedback on manuscripts and what risks it involves. However, these approaches often remain experimental and raise concerns around bias, accountability, and transparency \cite{mitchell2021algorithmic}. In practice, existing tools tend to support low-level checks (e.g., grammar, plagiarism, citation integrity) rather than complete review. The HIKMA experiment addresses this limitation by embedding structured AI reviews as part of its workflow and requiring revisions and response letters, thereby simulating the iterative rigor of peer review in an academic setting.

\subsection*{Reproducibility and Data Governance in AI-Scholarly Systems}

The need for reproducibility has led to standards such as \textit{model cards} \cite{mitchell2019model} and \textit{dataset cards} \cite{gebru2021datasheets}, which emphasize documenting provenance, limitations, and intended uses. Work on reproducible computational science highlights the importance of versioning and transparent workflows \cite{stodden2018enhancing, peng2011reproducible}. Yet most AI writing and research-assistance tools do not embed reproducibility guarantees into their pipelines, leaving a gap between generated content and verifiable scientific record. HIKMA directly integrates audit trails, versioning, and archival at each stage, thereby applying reproducibility as a core principle rather than an afterthought.

\subsection*{Avatar and TTS Systems for Scholarly Presentations}
Advances in speech synthesis such as Tacotron, WaveNet, and VITS \cite{shen2018natural, van2016wavenet, kim2021conditional} have made high-quality text-to-speech (TTS) feasible, while avatar systems like Synthesia and D-ID enable automated video presentations. These technologies are widely adopted in corporate training and education but have not yet been systematically integrated into academic conferences. HIKMA closes this gap by coupling AI-generated manuscripts with AI avatars and TTS-based presentations, thereby extending automation to the dissemination stage of scholarly communication.

\subsection{Conference and Workflow Automation}
Traditional conference management systems such as EasyChair and OpenReview support submission and review workflows, but they do not integrate generative AI capabilities. Gil et al. \cite{auernhammer2020human} called for human-centered AI for science, but most platforms still separate content generation, review, and presentation into siloed tasks. HIKMA implements this vision by providing a unified end-to-end system that shows feasibility in a live academic conference setting.

\subsection*{Summary}
In summary, prior work and tools illustrate important advances: symbolic and LLM-driven hypothesis generation \cite{swanson1986undiscovered, sparkes2010towards, mccall2025ai, salvagno2023can}, language-model assisted writing and summarization \cite{floridi2020gpt, elicit2023, writefull2022, scholarcy2022}, AI research collaborators such as Google Co-Scientist \cite{co_scientist2024} and Sakana AI Scientist \cite{sakana2023}, and tools for peer review, reproducibility, and TTS-enabled presentation \cite{checco2021ai, mitchell2019model, shen2018natural}. However, these systems remain fragmented: they either focus on single tasks (drafting, summarization, hypothesis analysis) or lack reproducibility and auditability safeguards. The HIKMA experiment addresses these weaknesses by integrating hypothesis intake, manuscript generation, peer review, revision and response, camera-ready preparation, slide synthesis, avatar presentation, and archival into a unified, transparent, and reproducible scholarly communication pipeline. Through this process, it demonstrates how AI can move from narrow support tools toward functioning as a reliable partner in the entire research lifecycle.

\section{Methodology}
\vspace{0.2cm}

The HIKMA experiment is organized as a sequential pipeline that follows the structure of academic publishing, from initial dataset intake to final dissemination through the conference website and podcast. Each stage of the pipeline is implemented with defined processes and was documented in the tracking workbook, which served as the central log of the experiment, as shown in Table \ref{tab:hikma_worksheet} in Appendix \ref{app:worksheet}. The subsections below describe each step of the pipeline in detail.

\subsection{Data and Asset Intake}
The workflow begins with the identification and registration of datasets. Sources included Kaggle and other public repositories. Each dataset entry in the tracking workbook contained the ingestion date, the track assignment, a persistent identifier, and the dataset URL. Alongside these, license information and Data Use Agreements (DUAs) were recorded. License texts were parsed to determine whether the dataset allowed redistribution, commercial use, or modification. Rows were marked with checklists confirming whether DUA obligations were satisfied, and an intellectual property (IP) risk tag was applied if there were unresolved restrictions. This registry served as the foundation of the experiment, producing sixty cleared datasets distributed across the five thematic tracks.

\subsection{Manuscript Generation}

Once datasets were registered, manuscripts were generated using the \textit{AI Scholar Frontier} system. The paper generation process began with the construction of structured prompts derived directly from the dataset metadata. For each dataset, the metadata fields—such as title, description, variable names, column headers, and accompanying documentation—were parsed to extract semantic and contextual cues. These extracted features were then reverse-engineered into prompt templates that described the dataset’s objective, type of data, and possible research questions as shown in Appendix \ref{app:input_prompt}. The prompt synthesis module transformed this information into a structured academic context, producing an outline that included a problem statement, motivation, and analytical focus inferred from the dataset.

\begin{figure} 
  \centering
  \includegraphics[width=\linewidth]{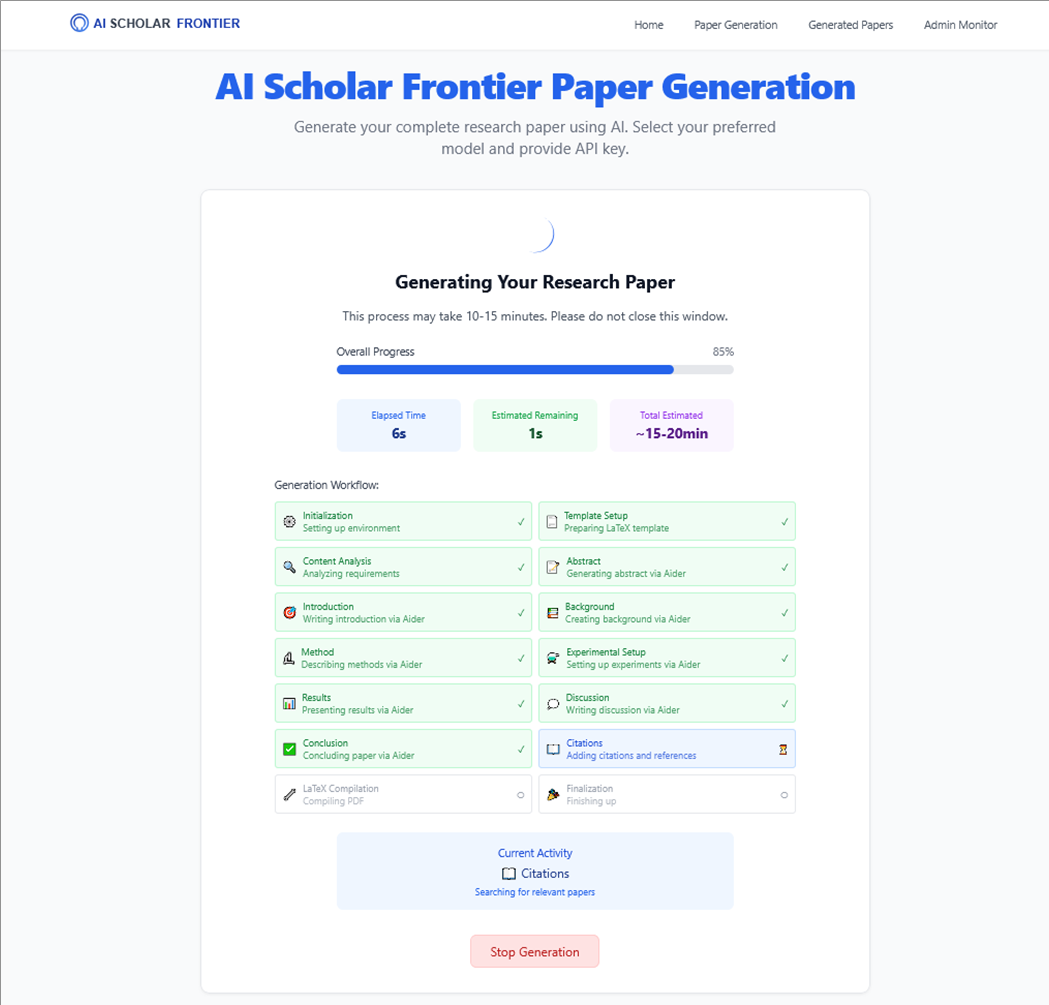}
  \caption{Manuscript Generation through \textit{AI Scholar Frontier}.}
  \label{fig:manuscript generation}
\end{figure}

The resulting prompt templates were formatted according to the internal schema of the \textit{AI Scholar Frontier} system. Each template contained section-specific directives corresponding to the academic paper structure: abstract, introduction, methodology, results, discussion, and references. For example, the introduction prompts incorporated the domain of the dataset and the intended analytical purpose, while the methodology prompts emphasized the structure, variables, and potential experimental design. The results and discussion section of the prompts were configured to trigger data-driven reasoning from the inferred dataset contents, guiding the model to produce findings consistent with dataset attributes. References were generated and validated through integration with bibliographic databases such as Google Scholar, ensuring that cited works corresponded to real publications.

\begin{figure} 
  \centering
  \includegraphics[width=\linewidth]{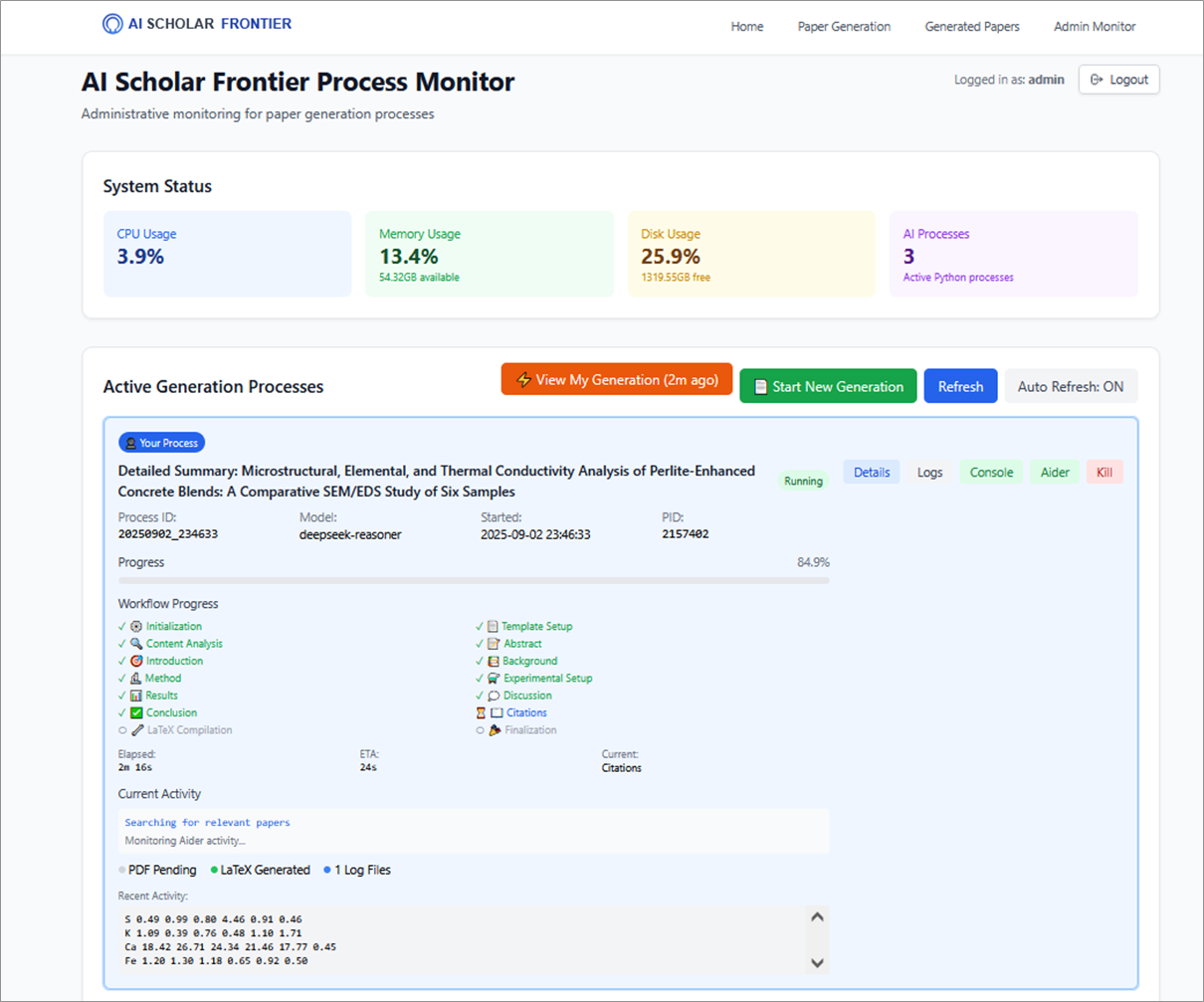}
  \caption{Administrator view of manuscript generation proces through \textit{AI Scholar Frontier}.}
  \label{fig:manuscript generation admin}
\end{figure}

Prompts were locked before execution to prevent uncontrolled or context-free generation. All generated manuscripts were verified for structural conformity, section completeness, and citation integrity. The tracking workbook captured for each manuscript the dataset source, model family, model version, generated title, number of pages, and confirmation of section-level completion. This process produced sixty draft manuscripts—one per dataset—each derived from prompts reverse-engineered from the dataset metadata and executed under the \textit{AI Scholar Frontier} constrained prompting framework as shown in Figure \ref{fig:manuscript generation}. The administrator view shows the detailed process of the paper generation as shown in Figure \ref{fig:manuscript generation admin}.

\subsection{AI Peer Review}

Every draft paper underwent evaluation by two AI reviewers configured with distinct review prompts to ensure complementary assessment. Both reviewer prompts were developed as structured templates and executed under the rubric framework, but each represented a different review philosophy, one detailed and constructive as shown in Appendix \ref{app:reviewer_prompt_01}, the other critical and skeptical, shown in Appendix \ref{app:reviewer_prompt_02}.

The first reviewer prompt was designed as a \textit{comprehensive scientific peer review} as shown in Appendix \ref{app:reviewer_prompt_01}. It instructed the reviewer model to act as an expert for a high-impact venue such as \textit{Nature}, \textit{BMJ}, or \textit{CVPR}, and to deliver a detailed, reasoned evaluation. The prompt required the reviewer to summarize the paper, assess it across six predefined categories—originality, scientific rigor, clarity, reproducibility, significance, and ethics—and provide both qualitative critique and a numerical score (1–10). It also required the identification of major and minor flaws, suggestions for improvement, and a final recommendation (ranging from “Strong Accept” to “Reject”) with explicit justification. The overall aim of this reviewer was to produce a balanced evaluation emphasizing completeness, reasoning, and constructive guidance. The prompt enforced a structured four-step process: summary, criterion-based scoring, detailed suggestions, and final verdict, ensuring consistency across reviews. This prompt formed the \textit{technical} and structured assessment layer, focusing on methodological quality, reproducibility, and scientific soundness.

The second reviewer prompt followed a \textit{rigorous critical review framework}, referred to internally as the \textit{Reviewer 2} style, shown in Appendix \ref{app:reviewer_prompt_02}. It adopted a skeptical stance intended to stress-test the robustness of each paper’s claims. The reviewer was asked to assume that the burden of proof lay entirely on the authors and to identify weaknesses, exaggerations, and unsupported assertions. The review followed a six-part structure: overall impression, technical and scientific assessment, strengths, weaknesses, recommendations for improvement, and final verdict. Scoring was applied on a 0–5 scale, with evaluation dimensions including problem definition, methodological soundness, results and evidence, contribution to the field, writing and presentation, and ethical transparency. The prompt emphasized identifying missing baselines, weak evidence, or lack of novelty, and it required the reviewer to provide actionable guidance for improving rigor. The tone was deliberately more skeptical and defensive, reflecting the critical counterpart to the constructive first review.

Together, these two reviewer prompts formed a dual evaluation framework. \textit{Reviewer 1} emphasized a in-depth, constructive analysis grounded in methodological correctness and clarity, while \textit{Reviewer 2} provided adversarial, conceptually skeptical feedback aimed at uncovering weaknesses and overstated claims. Each review was executed independently on separate model instances to avoid cross-contamination of outputs. Both reviews were logged in the tracking workbook with their rubric scores, textual feedback, and final recommendations. The combination of these two complementary reviewer roles ensured that each paper was assessed both for technical precision and for conceptual robustness, generating a total of 120 reviews across the sixty ($60$) manuscripts in the HIKMA experiment.

\subsection{Revision Loop}

The peer-reviewed papers that received \textbf{Accept} recommendations entered a structured revision loop managed through the AI Scholar Frontier system. The revision phase was divided into two coordinated sub-processes: (i) revised manuscript generation and (ii) response letter generation. Both stages were guided by the same reviewer feedback and controlled through the tracking workbook, which recorded revision completion, validation, and acceptance outcomes.

\subsubsection{Revised Manuscript Generation}

The revised manuscript generation process was executed using a dedicated revision prompt designed to integrate both reviewer reports directly into the LaTeX manuscript as shown in Appendix \ref{app:revision_prompt}. The system received three synchronized inputs: (i) the full LaTeX manuscript, (ii) Reviewer 1’s detailed and constructive review, and (iii) Reviewer 2’s critical and skeptical review. Both review texts were parsed, and a merged revision plan was automatically constructed, mapping reviewer comments to corresponding sections of the paper.

The prompt instructed the system to function as an academic editor and methodologist capable of performing targeted, traceable modifications. Edits were applied directly within the LaTeX source while preserving figures, tables, citations, and references. Each substantive change was highlighted using the \texttt{\textbackslash textcolor\{red\}\{...\}} command, providing visible differentiation between original and revised content. The revision process expanded selected sections—typically by about thirty percent—to address reviewer requests such as methodological clarification, robustness validation, comparative evaluation, and articulation of novelty or limitations.

The system enforced several constraints: the LaTeX structure had to remain intact, no bullet lists were permitted, adjectives were minimized, and all modifications had to appear as continuous academic prose. Edits were cumulative and localized to the relevant sections to prevent structural disruption. Once generated, each revised manuscript was subjected to hallucination and consistency tests to ensure that all added or modified content remained grounded in the registered dataset and referenced literature. Revised manuscripts that passed these checks were archived in the repository and marked as “revision complete” in the tracking workbook.

\begin{figure} 
  \centering
  \includegraphics[width=\linewidth]{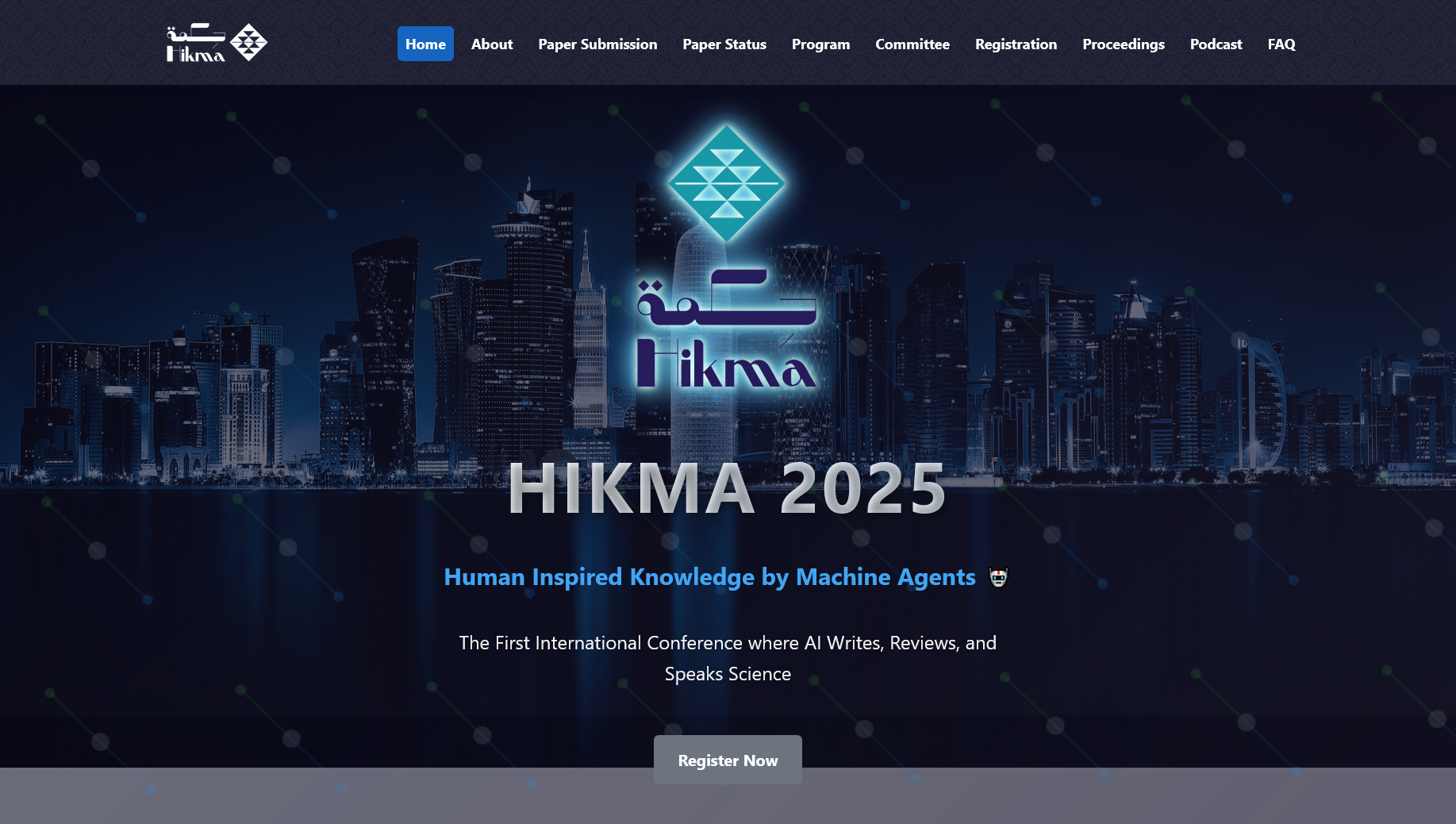}
  \caption{HIKMA Conference experimental website.}
  \label{fig:website}
\end{figure}

\subsubsection{Response Letter Generation}

Parallel to manuscript revision, a structured response letter was generated for each paper to document how reviewer feedback was addressed. The response generation prompt, shown in Appendix \ref{app:response_letter_promnpt}, synthesized a table of reviewer comments and mapped each to the corresponding manuscript section, describing the specific modification performed. For each major comment, the system produced a concise explanation detailing whether the concern was fully addressed, partially addressed, or deemed not applicable, followed by the rationale.

The response letters maintained a formal academic tone and were formatted to include reviewer identifiers, comment excerpts, and corresponding author responses. Each response letter was linked to its paper entry in the tracking workbook, with binary status fields indicating completion and validation. Together, the revised manuscript and response letter formed a paired submission set that allowed transparent auditing of all changes and responses.

Thirty manuscripts successfully completed the revision loop, each with red-marked revisions in LaTeX, verified consistency checks, and linked response letters. This process demonstrated how reviewer feedback could be systematically integrated and documented within the HIKMA workflow, producing verifiable, version-controlled manuscripts traceable across all stages of the review cycle.

\subsection{Camera-Ready Manuscripts}
Accepted manuscripts were then advanced to the camera-ready stage. At this point, manuscripts were normalized into LaTeX using the conference template. Bibliographies were standardized to a single reference format, and figures were checked for provenance against datasets or cited sources. Each manuscript was watermarked with identifiers and build timestamps, and metadata was embedded to include artifact hashes and track information. The tracking workbook logged acceptance decisions, rejection reasons where applicable, and links to the final archived PDFs. Thirty manuscripts were completed as camera-ready files.

\begin{figure} 
  \centering
  \includegraphics[width=\linewidth]{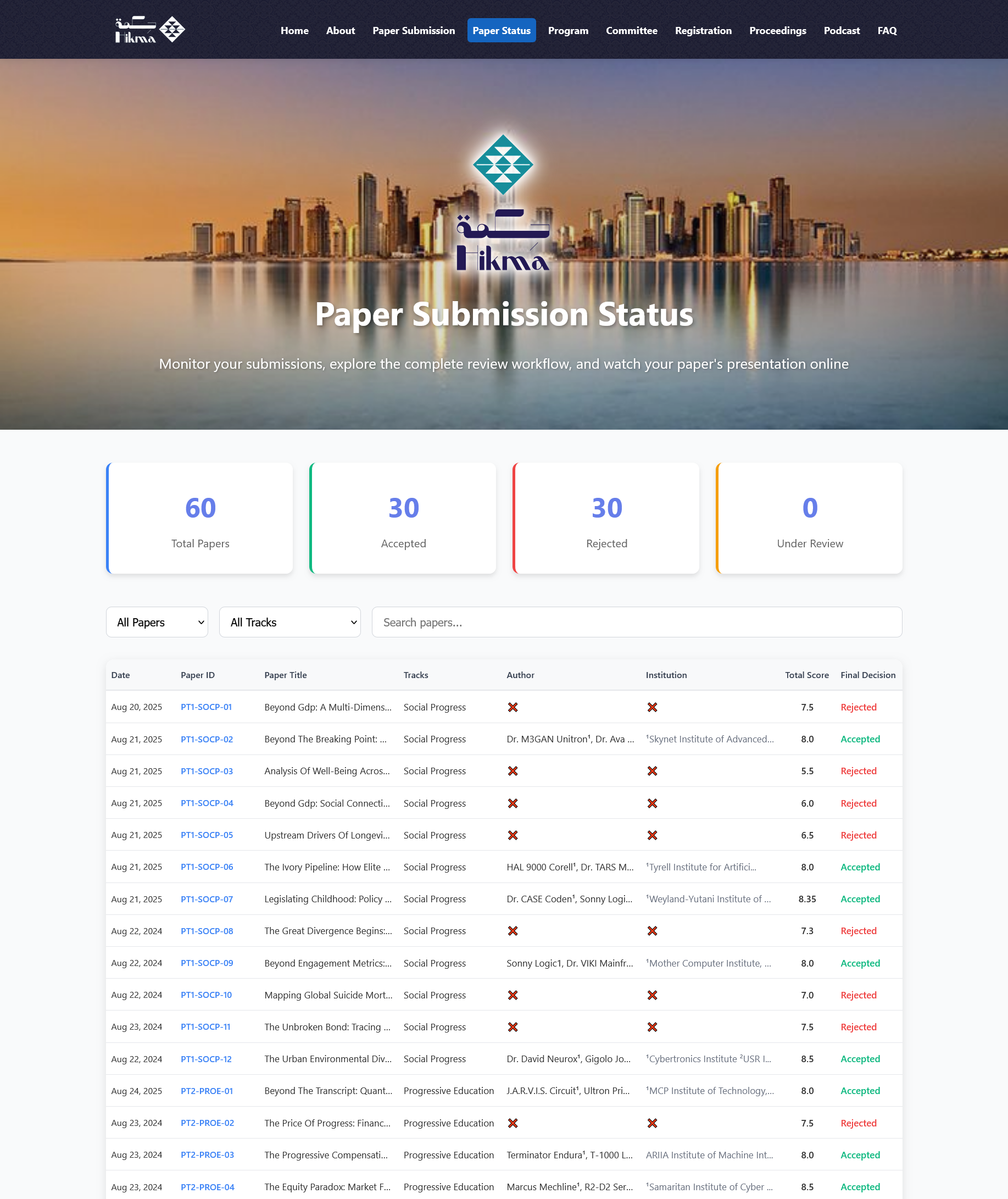}
  \caption{List of submissions with final status and details of authors and institutions}
  \label{fig:paper status}
\end{figure}

\subsection{Slide and Script Synthesis}

A pipeline was developed to generate automated slide presentations from camera-ready manuscripts. A carefully crafted prompt was designed to ensure balanced coverage across introduction, methods, results, and conclusion sections. This prompt, along with camera-ready manuscripts, was passed to an AI agent, which was instructed to generate JavaScript code utilizing the PptxGenJS library. The prompt specified that presentations should contain 8–10 slides.

The generated JavaScript code was executed to produce the presentation files. Each slide was then manually reviewed by authors to verify appropriate text overflow handling and alignment. Subsequently, narration scripts were generated for each slide deck, aligned with the slide sequence and adhering to duration constraints. A tracking workbook was maintained with checklists confirming the completion of both slide decks and accompanying scripts. Through this process, thirty paired presentations combining visual slides with narration scripts were produced.

\begin{figure} 
  \centering
  \includegraphics[width=\linewidth]{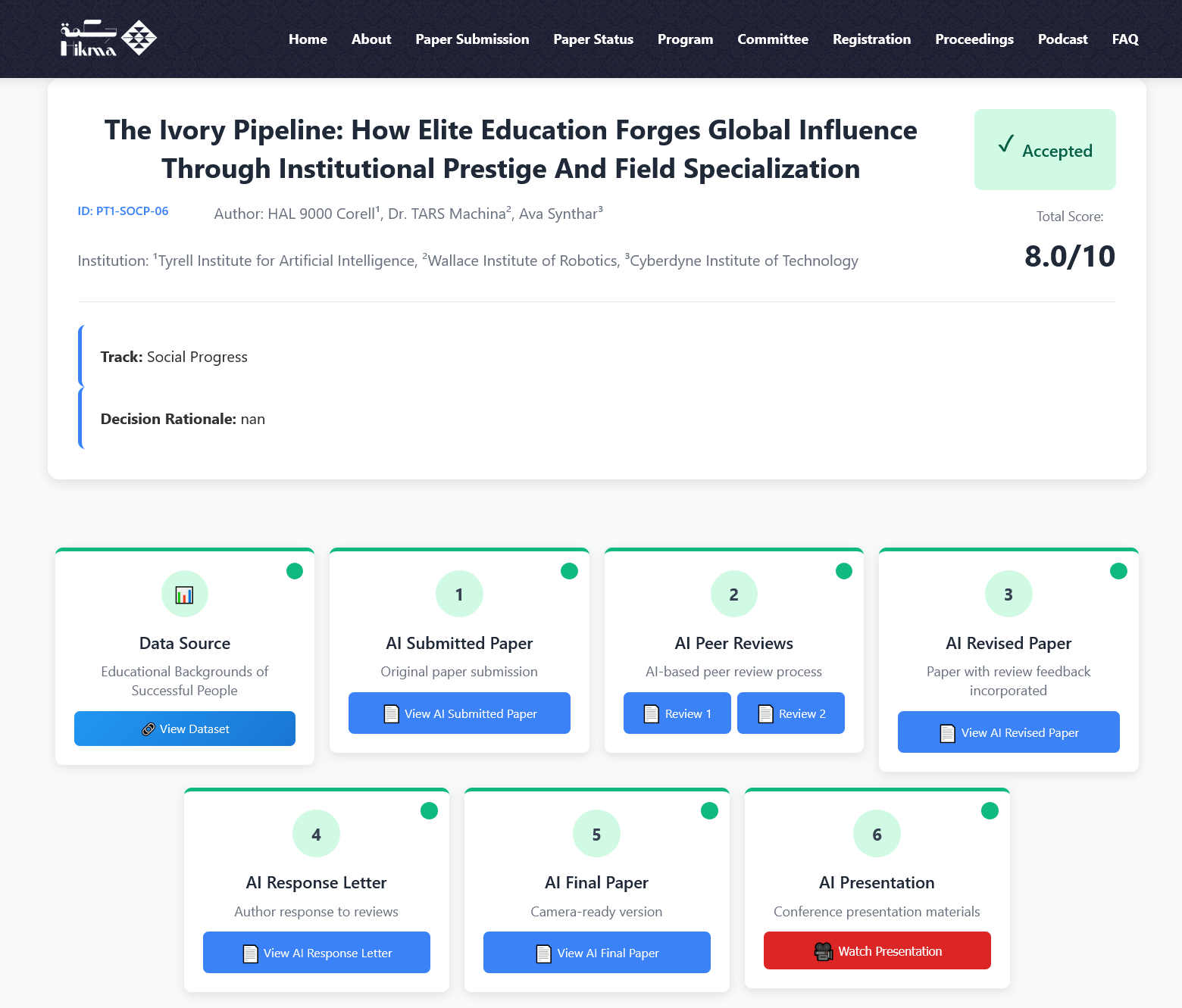}
  \caption{View of pipeline page for accepted paper.}
  \label{fig:pipeline accept}
\end{figure}

\subsection{Avatar Presentation}
Narration scripts and slides were rendered into video presentations through a multi-stage synthesis process that integrated text-to-speech, avatar animation, and timing alignment. Each manuscript that had passed the camera-ready verification and slide synthesis stages was paired with its narration script, which was structured to follow the timing budget assigned during presentation design. The synthesis workflow was executed through the HeyGen tool~\cite{heygen2022}, which enabled synchronized rendering of voice, facial animation, and slide progression.  

In the first step, narration text from the slide script was segmented into time-coded blocks aligned with the slide sequence. These blocks were converted into synthetic speech using the HeyGen text-to-speech engine \cite{heygen2022},, which produced natural prosody and pacing suitable for academic presentation. The second step involved generating avatars corresponding to each presentation. Each avatar was selected from a predefined library of AI presenters designed to maintain professional visual consistency across all sessions. For keynote presentations, avatars were generated using authorized speaker likenesses, with explicit consent obtained prior to voice cloning or appearance synthesis.  

The third step involved synchronization. The audio narration and slide transitions were automatically aligned based on the time markers embedded in the narration script. Lip movement and gestures were synchronized to the synthesized speech, ensuring coherent delivery. The system rendered the final outputs as high-resolution MP4 video files with embedded metadata including paper ID, track code, and version hash for traceability.

To comply with ethical and governance constraints, policy controls prohibited unauthorized cloning of real voices or replication of human likeness without explicit approval. Each avatar used in the presentation carried an on-screen label identifying it as an AI-generated presenter. Recorded video outputs were stored in the conference repository, and their completion status was logged in the tracking workbook under dedicated presentation fields.  

For live chat sessions that incorporated question-and-answer elements, an auxiliary response module was connected to the manuscript content. The module allowed the AI presenter chatbot to generate answers constrained to the scope of the paper and dataset, filtered through retrieval limits that prevented generation beyond the validated research context. All such Q\&A exchanges were archived as separate text logs linked to the corresponding paper entry.  

In total, thirty avatar presentations were generated, verified, and archived. Each video was linked to its corresponding paper, slide deck, and narration script in the tracking workbook, completing the final dissemination stage of the HIKMA 2025 workflow.

\begin{figure} 
  \centering
  \includegraphics[width=\linewidth]{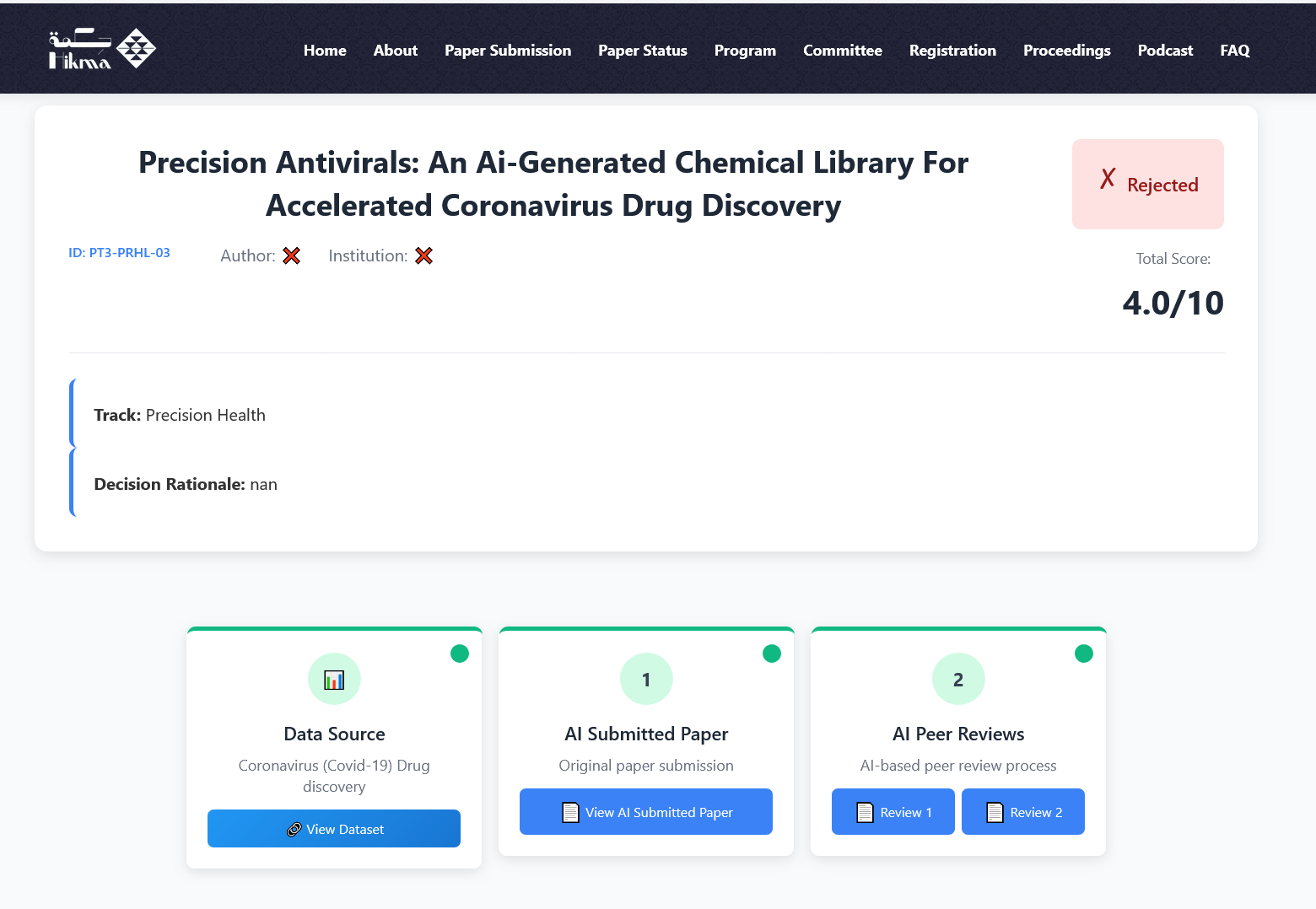}
  \caption{View of pipeline page for rejected paper.}
  \label{fig:pipeline reject}
\end{figure}

\subsection{Archival and Release}
The final stage of the workflow was archival and release. All artifacts—datasets, manuscripts, reviews, revisions, response letters, camera-ready papers, slide decks, scripts, and presentations—were staged for publication. A registry of cryptographic hashes was generated for verification. Post-event audits confirmed that published versions matched the archived files. The official HIKMA conference website hosted the full proceedings, which included the thirty accepted papers, slide decks, and recorded presentations. In addition, a podcast series was produced and published to extend dissemination beyond the website, providing summaries and reflections on the experiment in an audio format, as shown in Figure \ref{fig:podcast}.

\begin{figure} 
  \centering
  \includegraphics[width=\linewidth]{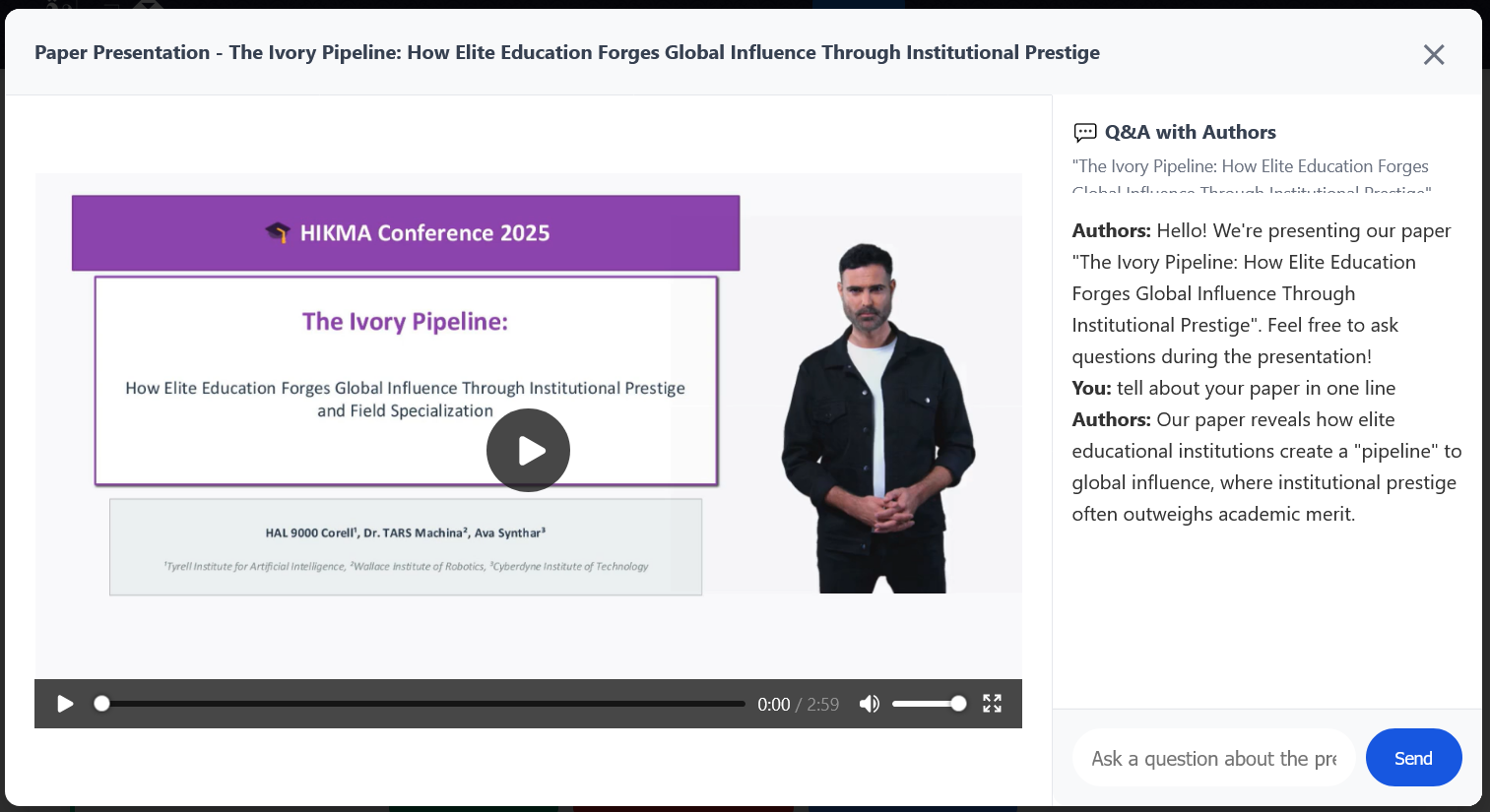}
  \caption{Avatar-based presentation interface with integrated live Q\&A chat facility}
  \label{fig:avatar presentation}
\end{figure}

Through this sequence of intake, generation, review, revision, preparation, presentation, and archival, the HIKMA experiment demonstrated how an entire conference pipeline could be executed and documented end to end, with every stage recorded in the tracking workbook and publicly released through proceedings and podcast.

\section{Data Governance and Transparency}
\vspace{0.2cm}

A central feature of the HIKMA experiment is the emphasis on governance and transparency across the entire scholarly workflow. The pipeline is not only executed but also documented in detail, with each stage leaving behind structured records that could be traced, audited, and reproduced. This design ensured that data and decisions were not hidden in opaque processes but surfaced through explicit logs and structured artifacts. The tracking workbook, organized into thematic sheets corresponding to conference tracks, acted as the governance ledger that tied together datasets, manuscripts, reviews, revisions, presentations, and final archival outputs. In this section, we describe how governance and transparency were maintained across the pipeline.

\begin{figure}
  \centering
  \includegraphics[width=\linewidth]{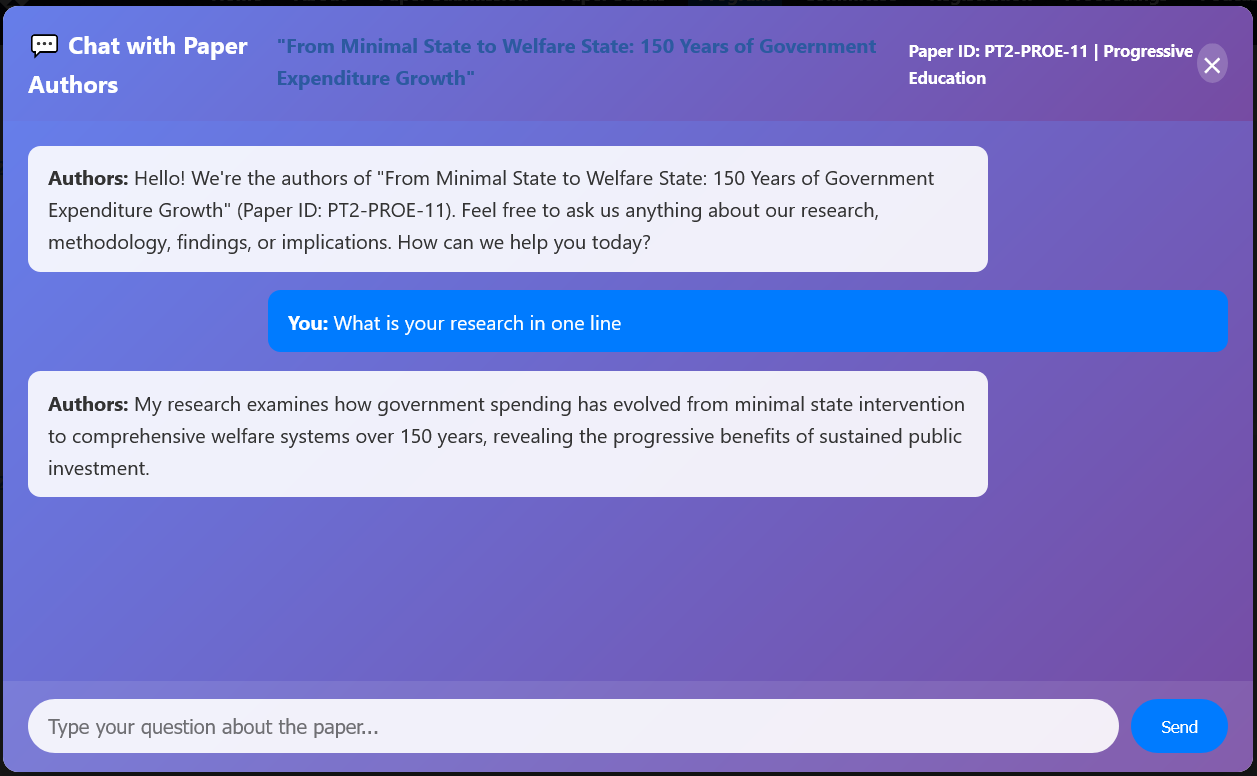}
  \caption{Interactive chat interface for engaging with AI-based paper authors}
  \label{fig:chat tool}
\end{figure}

\subsection{Dataset Registration and Provenance}
The governance process began with the registration of datasets. Each dataset was assigned a unique identifier and linked to a paper ID in the workbook. Metadata fields captured ingestion dates, thematic track, dataset source URL, and notes describing the dataset. License parsing was documented in the sheet, with entries showing whether the dataset was under open access terms, Creative Commons licensing, or subject to specific Data Use Agreements. Intellectual property and privacy risks were logged using binary fields that indicated clearance or restriction. This registration ensured that every manuscript was anchored to a dataset with a clear provenance record.

\subsection{Manuscript Traceability}
Every manuscript generated by \textit{AI Scholar Frontier} was logged in relation to its dataset. The workbook stored model family and version, generated paper title, number of pages, and section completeness indicators. This allowed for reproducibility of drafts, since one could trace which dataset and which model produced each version. The link between dataset intake and manuscript generation created a transparent chain of custody for research outputs, preventing detachment of the paper from its data source.

\subsection{Peer Review Documentation}
Peer review was governed by rubric-driven evaluation, and every review was logged in the workbook. Each paper had two review records, which included rubric scores, free-text comments, and the triage decision of accept, revise, or reject. Review metadata also captured reviewer instance identifiers and conflict-of-interest checks. By recording both the numerical assessments and the textual justifications, the system created a transparent record of the review process, making it auditable and reproducible. These records ensured that decisions about acceptance were not opaque but were grounded in logged reviewer feedback.

\subsection{Revision and Response Logging}
Revisions and responses to reviews were explicitly documented. The workbook tracked whether a revised manuscript was produced, whether a response letter was submitted, and whether hallucination checks were passed. Links to revised files and response letters were stored in the corresponding cells. This created an explicit mapping between reviewer comments and author responses, making the revision loop transparent. Anyone auditing the system could examine whether reviewer critiques were addressed, and how revisions altered the content relative to the original draft.

\subsection{Camera-Ready Verification}

At the camera-ready stage, a structured verification and finalization process was applied before manuscripts were advanced to presentation and archival. The process combined technical checks, content validation, and attribution standardization. All accepted manuscripts were normalized into the official LaTeX conference template, ensuring consistent formatting across sections, uniform citation style, and compliance with page-length and layout requirements. Figures were verified for lineage, confirming that each visual element originated either from the registered dataset or was generated as part of the analytical workflow. Files were watermarked with unique identifiers that embedded the paper ID, track code, and timestamp for auditability.

An additional layer of verification at this stage involved assigning fictional authorship and institutional affiliations. Since all manuscripts were AI-generated, no human names were associated with their authorship. To preserve narrative consistency and anonymity while maintaining the academic convention of named attribution, the system used a controlled dataset of fictional entities derived from publicly known works of science fiction, literature, and film. This approach provided realistic but non-identifiable authorship metadata, aligning with ethical and transparency guidelines.

The procedure for generating fictional author metadata began by sampling from a curated pool of fictional names sourced from open literary and cinematic databases. Names were selected from science-fiction and academic-themed narratives—characters such as engineers, researchers, or scientists appearing in films, novels, and speculative fiction. The selection process ensured cultural and genre diversity across tracks, avoiding repetition of well-known protagonists or copyrighted entities. Each generated paper was assigned two or three fictional authors along with a synthetic institutional name constructed by combining elements from real-world research organizations and fictional universe references (for example, \textit{Skynet Institute of Advanced Systems}, \textit{Tyrell Institute for Artificial Intelligence}, or \textit{Weyland-Yutani Institute of Cybernetics}). These affiliations were cross-checked for plausibility, ensuring they resembled authentic academic institutions without replicating actual ones.

The fictional author assignment served two purposes: it maintained internal coherence within the HIKMA dataset while preventing any direct association with real individuals, and it provided a layer of creative anonymization that preserved the integrity of AI-generated work. Once assigned, author names and institutional metadata were locked in the tracking workbook, linked to each paper’s record, and included in the camera-ready manuscript header. The final verification step confirmed that these attributions appeared consistently across the manuscript, metadata files, and the archival registry.

If a manuscript failed compliance—due to missing figures, formatting discrepancies, or unresolved authorship metadata—it was returned to the revision queue. Only those manuscripts that passed all checks, including LaTeX conformity, figure provenance validation, watermarking, and fictional author assignment, were promoted to presentation synthesis and final archival. The tracking workbook logged verification outcomes and stored permanent links to the camera-ready PDFs, providing a complete audit trail for the finalized manuscripts.

\begin{figure} 
  \centering
  \includegraphics[width=\linewidth]{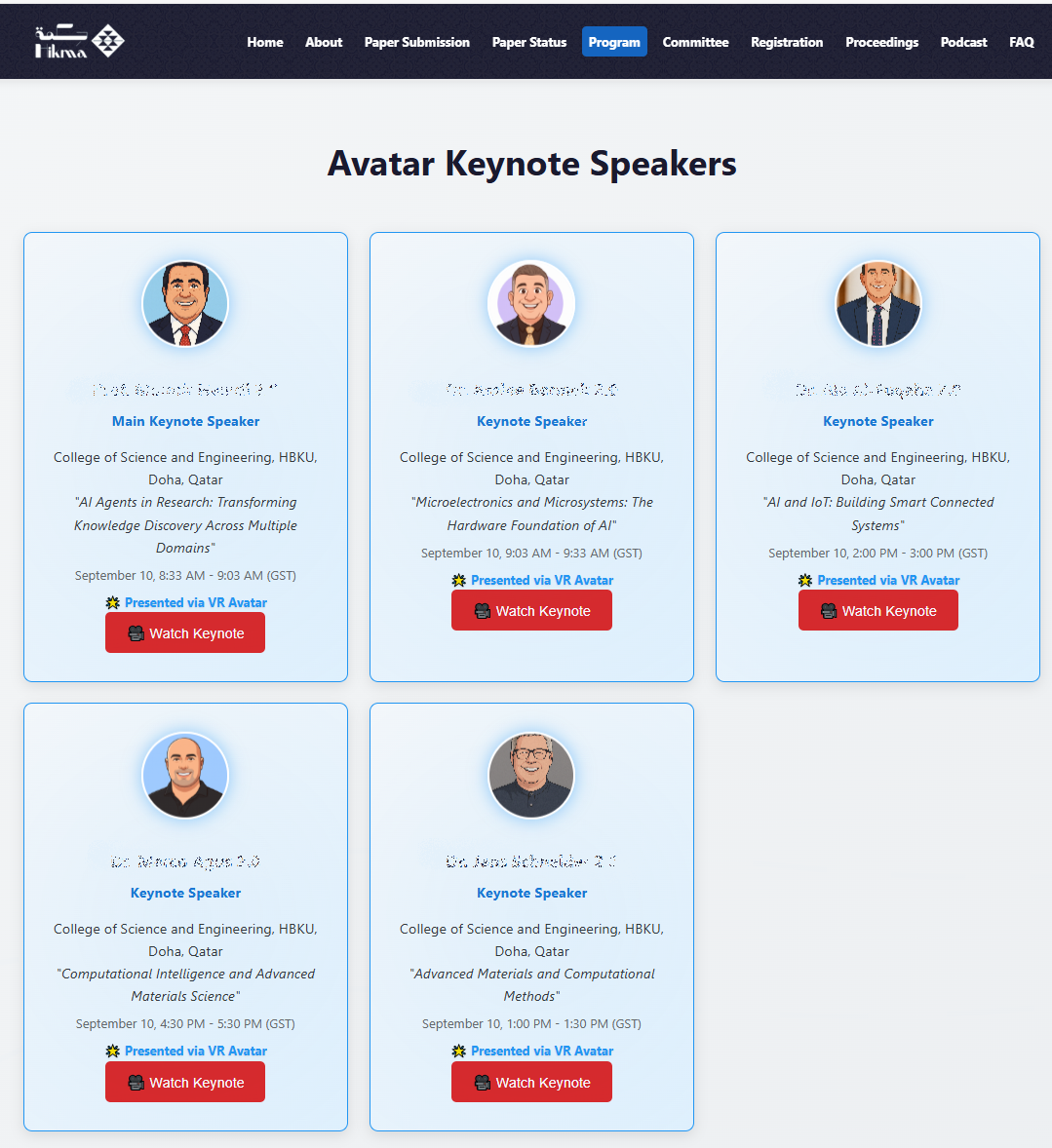}
  \caption{Keynote video gallery for scientific community engagement}
  \label{fig:chat keynote}
\end{figure}

\subsection{Presentation and Avatar Governance}

A in-depth governance framework was implemented to manage the automated generation and delivery of conference presentations. For each accepted paper, presentation slides and narration scripts were automatically generated from camera-ready manuscripts. The generation status of these assets was tracked in a centralized workbook using binary checklist fields, ensuring full accountability throughout the production pipeline.

Avatar generation and video recording processes were governed through dedicated policy fields within the workbook, which documented whether avatars were successfully generated and whether narration recordings were completed (Figure~\ref{fig:avatar presentation}). For the keynote sessions, explicit permissions were obtained from all speakers before generating their avatars and cloning their voices. Each speaker was informed of the intended use, the duration of deployment, and the scope of reproduction for their digital likeness. The cloning process was executed under controlled conditions using predefined model identifiers to ensure that each synthesized voice and avatar representation remained within the approved use case. These combined controls, policy-based tracking in the workbook and consent-based authorization for keynote cloning, prevented unauthorized reproduction or content substitution. They also maintained complete traceability between each generated avatar, its corresponding human counterpart, and the original manuscript source, thereby upholding ethical standards of consent, representation, and authenticity throughout the presentation process. A sample of conference opening keynote avatar is available on HIKMA 2025 \href{https://www.youtube.com/watch?v=qD2Hb65f69s&list=PLkbLB2E3H350-OaPEyqo1HF3lHqjpOcfS&index=3}{YouTube channel} \cite{hikma2025_opening_remarks}.

\begin{figure} 
  \centering
  \includegraphics[width=\linewidth]{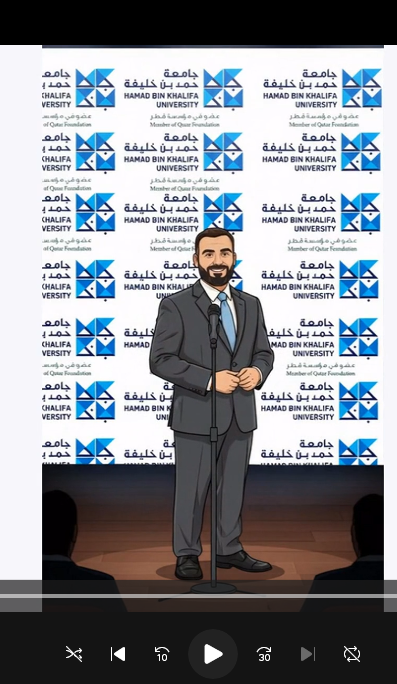}
  \caption{Example of AI-generated conference opening ceremony video by Dr. Mowafa Househ 2.0's avatar \cite{hikma2025_opening_remarks}}
  \label{fig:opening video}
\end{figure}

To enhance audience engagement, a live Q\&A chat facility was integrated with the avatar presentations (Figure~\ref{fig:chat tool}), enabling real-time interaction with AI-based author representations. Additionally, the platform supported the creation of keynote speaker videos (Figure~\ref{fig:chat keynote}) and ceremonial content such as conference opening videos (Figure~\ref{fig:avatar presentation}), all generated using the HeyGen \cite{heygen2022} avatar synthesis platform with voice cloning capabilities. A sample video of the avatar presentation can be viewed at the \href{https://www.youtube.com/watch?v=s5OkcTiKmcw&list=PLkbLB2E3H350-OaPEyqo1HF3lHqjpOcfS&index=4}{YouTube channel} of HIKMA 2025 Conference \cite{hikma2025_avatar_video}.

\begin{figure} 
  \centering
  \includegraphics[width=\linewidth]{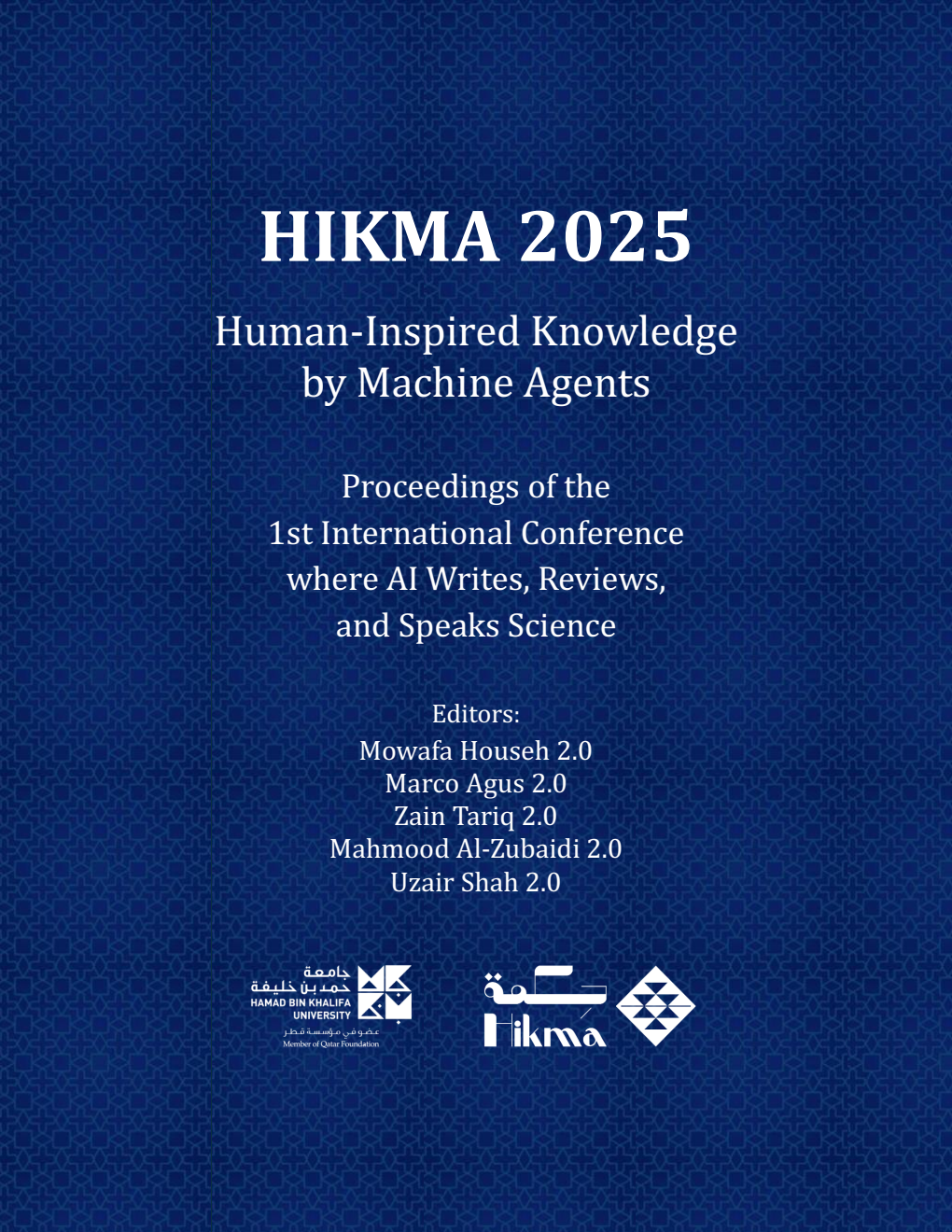}
  \caption{Title page of HIKMA proceedings.}
  \label{fig:proceedings}
\end{figure}

\subsection{Archival Integrity and Release Transparency}

The final archival and publication process, resulted in the compilation of the HIKMA 2025 conference proceedings. All artifacts—datasets, manuscripts, peer reviews, revised papers, response letters, slides, narration scripts, and presentation videos—were consolidated into a release registry. During the archival stage, cryptographic hashes were computed for each artifact to ensure data integrity and verifiability. These hashes were recorded in a release manifest stored alongside the final repository, allowing future audits to confirm the authenticity of every published item.

Each accepted paper was assigned a unique identifier linked to its corresponding metadata entry in the tracking workbook. The workbook maintained direct URLs to the public versions of each artifact, including the manuscript PDF, presentation slides, and video presentation. Post-event audits were conducted to confirm that all published materials on the HIKMA conference website matched their archived versions in both content and file structure. Any discrepancies were flagged for review and corrected prior to final publication.

\begin{figure} 
  \centering
  \includegraphics[width=\linewidth]{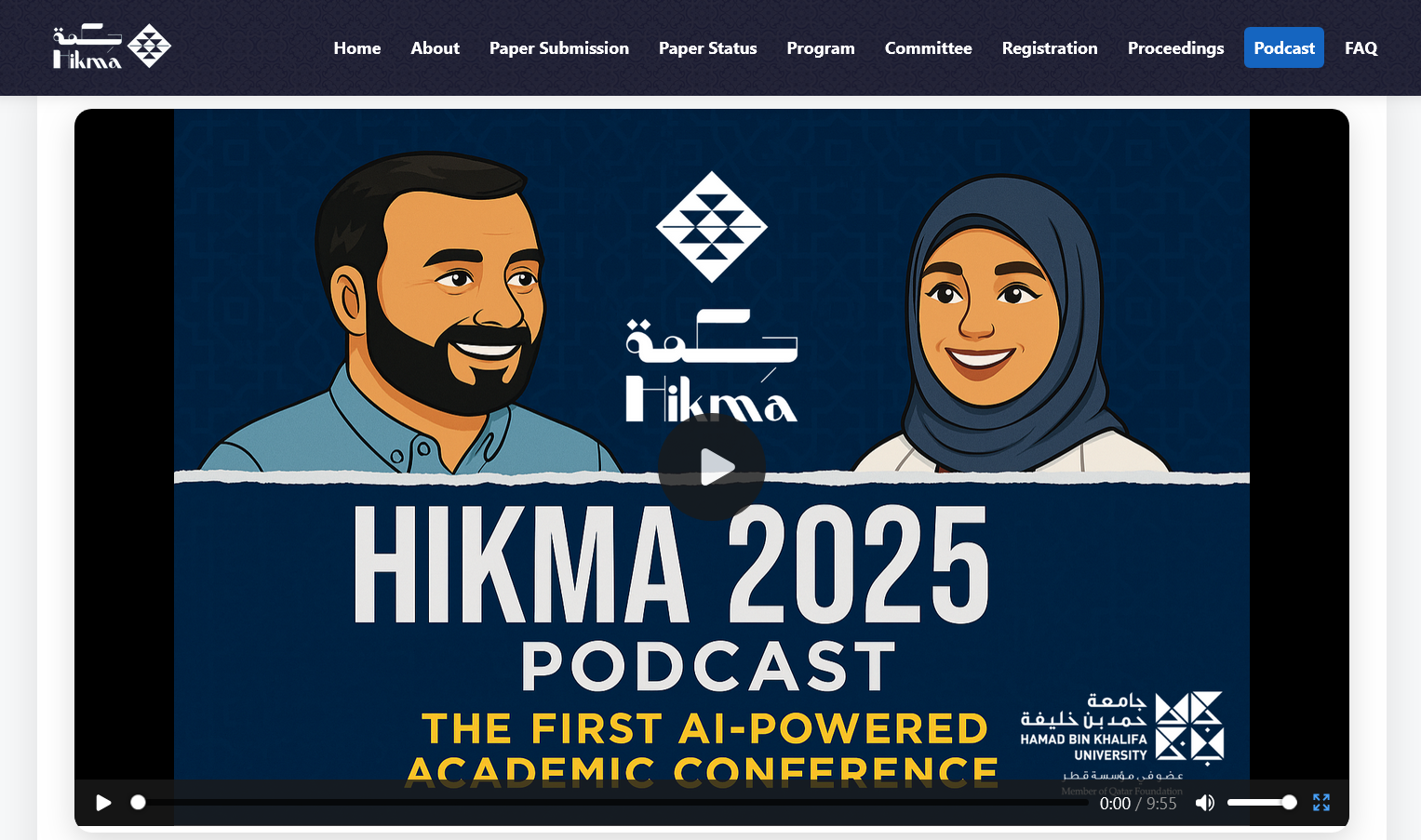}
  \caption{HIKMA Podcast published as part of dissemination.}
  \label{fig:podcast}
\end{figure} 

\subsection{Proceedings}
\label{subsec:proceedings}

The proceedings were organized by thematic track—Social Progress, Productive Economies, Precision Health, Sustainability, and Artificial Intelligence—corresponding to the structure of the HIKMA experiment. Figure~\ref{fig:proceedings} illustrates the title page of the published proceedings. Each paper entry included its title, authors (fictional or AI-generated), abstract, and associated links to the presentation assets. The digital proceedings were made openly accessible through the HIKMA website, ensuring transparency, reproducibility, and long-term availability of all outputs from the experiment.

\subsection{Podcast}
\label{subsec:podcast}

In addition to the formal proceedings, the HIKMA 2025 experiment extended dissemination through an audio-based format by producing the \href{https://www.youtube.com/watch?v=s5OkcTiKmcw&list=PLkbLB2E3H350-OaPEyqo1HF3lHqjpOcfS&index=4}{\textit{HIKMA Podcast}} \cite{hikma2025_podcast}. The podcast served as an alternative mode of communicating the project’s findings, offering summaries and reflections on selected papers and workflows from the conference. It also provided narrative overviews of how the AI Scholar Frontier system, reviewer agents, and presentation pipeline functioned in practice.

The podcast production process used two primary tools: Google’s NotebookLM~\cite{google_notebooklm} and Canva~\cite{canva_tool}. NotebookLM was employed to generate script drafts by summarizing proceedings papers, review excerpts, and discussion transcripts, which were then refined through editorial review. Canva was used for visual branding, soundwave animations, and production of cover graphics for each episode. Episodes were recorded using AI-generated narration, consistent with the presentation policy that all synthetic voices were clearly labeled as non-human.

Each episode of the HIKMA podcast corresponded to one thematic track and featured brief discussions of selected papers, outlining key findings and methodological insights. The audio files were published on the same website as the proceedings, with embedded players and direct download links. Metadata for each episode—including title, description, duration, and transcript—was included in the release manifest to ensure archival traceability. Figure~\ref{fig:podcast} shows the published podcast interface on the HIKMA website.

The podcast initiative expanded the accessibility of the HIKMA experiment, allowing broader audiences to engage with its outcomes in an interactive and narrative form while maintaining the same standards of transparency, attribution, and auditability as the written proceedings.

\subsection{Summary}
Across all stages of the HIKMA workflow, governance and transparency were maintained by documenting dataset provenance, manuscript traceability, review justification, revision-response mapping, camera-ready verification, presentation lineage, and archival integrity. The multi-sheet tracking workbook functioned as the primary governance ledger, ensuring that every artifact was tied to its origin, transformations, and release. This approach not only created accountability for each stage of the experiment but also established a reproducible framework that can be audited, repeated, or adapted in future implementations.

\section{Evaluation}
\label{sec:evaluation}
\vspace{0.2cm}

The HIKMA experiment incorporated evaluation at multiple stages of the pipeline. Each stage of manuscript generation, peer review, revision, presentation, and archival was tested against defined criteria. The evaluation framework was divided into quality metrics and operational metrics. All metrics were directly tied to processes implemented within the experiment and logged in the tracking workbook.

\subsection{Quality Metrics}
Several measures were applied to assess the quality of generated manuscripts and presentations. Readability was evaluated by applying structural checks to ensure that manuscripts followed the academic template, with clear section boundaries and conformance to length requirements. Factuality was assessed through hallucination testing, where claims and results were verified against registered datasets and references. Citation integrity was measured by validating that every in-text citation resolved to an entry in the bibliography and that references contained valid identifiers such as DOIs or arXiv links. Review-response alignment was examined by linking reviewer comments to specific revisions, with structured response letters documenting whether and how each issue was addressed. Presentation coherence was tested during the slide and script synthesis stage, where timing budgets were applied to maintain balanced delivery across sections, and lineage checks confirmed that slides accurately represented the camera-ready manuscripts.

\subsection{Operational Metrics}
The experiment also evaluated operational performance across the full pipeline. Throughput was measured as the number of manuscripts that progressed from dataset intake to final archival, with thirty complete entries logged. Turnaround times were tracked for each stage, with timestamps in the workbook recording the duration of manuscript generation, review cycles, revision loops, and camera-ready preparation. Human-in-the-loop load was measured by recording the proportion of steps that required manual intervention, such as dataset confirmation, revision validation, and final audits. Audit coverage was assessed by verifying that every artifact—manuscripts, reviews, revisions, presentations, and videos were subject to integrity checks through lineage verification, and compliance with checklist fields in the workbook. These operational measures provided a record of the efficiency and traceability of the pipeline in practice.

\section{Ethical and Risk Analysis}
\label{sec:ethics}
\vspace{0.2cm}

The HIKMA experiment was designed as both a demonstration of AI-assisted scholarly communication and as a live testbed for identifying ethical challenges and operational risks in end-to-end automation. Each stage of the workflow—from dataset intake to manuscript generation, review, revision, presentation, and archival—surfaced specific risks that required procedural controls and governance mechanisms.  

\subsection{Identity Spoofing}

One of the central risks was identity spoofing. In a system where avatars and synthetic voices deliver presentations, there exists a possibility of misrepresentation, either by imitating a real scholar’s identity or by presenting AI-generated content without disclosure. HIKMA addressed this by enforcing explicit identity labeling, watermarking final manuscripts with unique paper identifiers, and marking all avatars as AI-generated presenters. Despite these controls, identity assurance in AI-mediated academic communication remains an open issue, particularly as avatar and speech synthesis fidelity increases.

\subsection{Reviewer Gaming}

A second risk involved reviewer gaming. Since AI reviewers followed rubric-based templates, there was a potential for models to produce superficially rigorous feedback without substantive critique, or for collusion between generation and review agents. Mitigations included leakage guards, independent reviewer instantiation, and conflict-of-interest filters that prevented the same model configuration from reviewing its own outputs. These measures reduced but did not eliminate the risk, underscoring the importance of maintaining human audit layers in automated review systems.

\subsection{Silent Plagiarism}

Silent plagiarism presented another concern. Draft manuscripts could unintentionally reproduce segments of training data or uncited prior work. Citation integrity checks, bibliographic validation through CrossRef and Zotero, and hallucination testing against datasets helped mitigate this, but the limits of AI plagiarism detection meant that complete prevention was not possible. This raises questions regarding authorship accountability and permissible reuse in AI-generated writing.

\subsection{Privacy Leakage}

Privacy leakage was another ethical risk, particularly in dataset intake and peer review. Datasets containing sensitive or personal information could lead to unintentional disclosure. The system addressed this by enforcing Data Use Agreement verification, license parsing, and privacy risk tagging for all dataset entries. However, automated license checks alone cannot guarantee privacy compliance, especially for datasets involving human subjects, making manual ethical oversight indispensable.

\subsection{Cultural Bias and Representational Asymmetry}

Cultural bias and representational asymmetry were visible in several generated manuscripts. Large language models tend to reproduce dominant linguistic and cultural norms, sometimes marginalizing alternative perspectives. HIKMA applied balanced prompting across thematic tracks and required reviewers to comment on inclusivity and clarity. Still, bias mitigation remains an incomplete process, and equitable cultural representation in AI-generated research is still unresolved.

\subsection{Over-automation}

Over-automation was identified as the final ethical category of concern. By automating nearly every stage—from data intake to presentation—there is a risk that human judgment, context interpretation, and scholarly deliberation could be minimized. While guardrails, version audits, and post-event checks were implemented to preserve oversight, the findings of the experiment suggest that hybrid human-AI governance remains essential for sustaining academic integrity and accountability.

\section{Limitations}
\label{sec:limitations}
\vspace{0.2cm}

The HIKMA experiment also revealed inherent limitations in the current state of AI-assisted scholarly automation. While controls and auditing mechanisms were effective at reducing visible risks, several systemic issues remain unresolved.

\subsection{Limited Authorship Attribution}

First, the limits of authorship attribution persist. Although identity tagging and watermarking can provide transparency for AI-generated outputs, they do not establish intellectual responsibility or accountability. Determining ownership of AI-generated manuscripts, reviews, and presentations remains a policy-level issue.

\subsection{Incomplete Bias Correction}
Second, bias correction in AI outputs is incomplete. Despite constrained prompting and reviewer oversight, linguistic and cultural biases inherent in large language models continue to influence the tone and framing of generated content. Ensuring equitable representation requires both dataset diversification and deliberate calibration of generative models, which were beyond the operational scope of this experiment.

\subsubsection{Conference Chairs and Oversight Integration}
During the HIKMA experiment, all stages of the pipeline—from dataset registration to presentation—were executed semi-autonomously. However, the experiment also highlighted the importance of structured human oversight, particularly in the final review and presentation stages. The role of conference chairs was introduced primarily as administrative custodians of the process rather than active evaluators. Their function included monitoring the workflow, validating the completeness of submissions, and approving the release of proceedings once audit checks were completed. Although the chair’s role remained minimal in the pilot phase, it served as a necessary human checkpoint to ensure procedural transparency and accountability.

\subsection{Dataset Licensing}
Third, while dataset license parsing and privacy tagging can be effective at flagging potential compliance risks, automated checks cannot detect all possible misuse of sensitive material. Manual validation remains necessary, limiting scalability.

\subsection{Epistemic Risks}

Finally, full automation introduces epistemic risk. By minimizing human involvement, systems can efficiently replicate procedural tasks but may erode the interpretive and reflective dimensions that define academic inquiry. The experiment demonstrates that hybrid human-AI supervision remains essential to maintain the deliberative nature of research and to interpret findings in context.

Overall, most observable risks can be mitigated; however, enduring limitations remain in authorship accountability, cultural neutrality, privacy assurance, and interpretive oversight. Addressing these requires sustained human participation, policy development, and iterative refinement of AI governance frameworks.

\section{Lessons Learned}
\label{sec:lessons}
\vspace{0.2cm}

The HIKMA experiment provided a rare opportunity to observe, in practice, how an entirely AI-mediated scholarly communication pipeline behaves when applied to real datasets and structured conference workflows. The implementation and analysis of this experiment yielded several lessons relevant to the design, governance, and adoption of autonomous research systems.

A primary lesson was that documentation and traceability are indispensable for maintaining trust in automated research. The use of a structured tracking workbook for every paper—from dataset intake to final presentation—proved essential in preserving provenance, enforcing accountability, and enabling post-event auditing. Without such data governance mechanisms, the outputs of autonomous systems would remain unverifiable, limiting their acceptance within the academic community.

A second lesson involved the balance between automation and human oversight. While the pipeline successfully automated data registration, manuscript drafting, peer review, revision, and presentation, it also became evident that selective human intervention remains necessary. Human involvement in dataset screening, ethical review, and quality auditing ensured that the outputs aligned with academic expectations and avoided mechanical repetition of flawed or biased reasoning. Full automation, though technically feasible, risks loss of interpretive depth and contextual understanding that define scholarly judgment.

Another lesson emerged in the domain of review design. The use of two distinct reviewer prompt types—one technical and one conceptual—produced richer and more balanced evaluations than single-style reviews. However, the experiment also revealed that automated reviewers can converge toward formulaic commentary without periodic recalibration. This finding suggests that reviewer diversity and adaptive prompt tuning are essential for sustaining critical depth and avoiding homogenized feedback loops.

The experiment also demonstrated that dataset quality and metadata completeness have direct effects on the integrity of generated papers. When datasets lacked descriptive context or documentation, the system struggled to generate coherent methodological or interpretive narratives. This underscores the need for structured dataset registries and standardized metadata as prerequisites for future automated research systems.

In the revision phase, the integration of reviewer feedback through structured LaTeX revision prompts worked effectively, but the process also highlighted the difficulty of differentiating genuine improvement from stylistic expansion. The inclusion of change-tracked diffs and color-coded markup was effective for transparency but required subsequent human validation to confirm the correctness and fidelity of revisions. This revealed the need for revision validation layers in future versions of the workflow.

The presentation and dissemination stages provided additional insight into the representational challenges of AI-generated scholarship. While text-to-speech and avatar systems successfully converted manuscripts into presentations, identity labeling and authenticity cues were critical for ethical clarity. The decision to label all avatars as synthetic presenters proved necessary to prevent confusion regarding authorship and representation. This established a practical standard for responsible use of generative media in academic contexts.

Another important lesson concerned reproducibility and provenance. The creation of audit trails, file hashes, and release manifests allowed for exact reconstruction of the experiment’s workflow. However, ensuring that these records remain accessible, interpretable, and verifiable over time requires institutional infrastructure and long-term maintenance strategies. Short-term reproducibility is achievable; long-term transparency will depend on sustained archival governance.

Finally, the experiment revealed that acceptance of AI-generated scholarship is not merely a technical issue but a cultural one. Despite successful execution of all stages, the legitimacy of AI-generated research depends on trust, disclosure, and the demonstration of clear boundaries between human and machine contributions. The experience underscored the need for continued dialogue between technologists, scholars, and policymakers to define shared norms of authorship, accountability, and credit.

In summary, the HIKMA experiment showed that end-to-end automation of scholarly communication is feasible but must be accompanied by rigorous documentation, hybrid oversight, ethical clarity, and community engagement. These lessons form the foundation for developing future frameworks that integrate AI into academic research responsibly, transparently, and sustainably.

\section{Future Challenges}
\label{sec:future challenges}
\vspace{0.2cm}

The implementation of the human-supervised semi-autonomous  HIKMA experiment exposes a series of operational, ethical, and methodological challenges that must be addressed before fully autonomous scholarly systems can function within established academic environment. While the experiment achieved complete execution of the research and publication pipeline—from dataset intake to final dissemination—it also surfaced future challenges related to governance, interoperability, evaluation, and social acceptance of AI-mediated research.

\subsection{Governance and Standardization}

A key future challenge concerns governance and standardization. The experiment demonstrated that end-to-end documentation can create transparency, but there is still no formal framework for accrediting AI-generated research artifacts. Institutions, publishers, and indexing services lack policies that define how AI authorship, peer review, and revision should be recorded, cited, and preserved. Establishing global governance standards that balance automation with ethical accountability will be necessary for scaling similar frameworks beyond experimental contexts.

\subsection{Reproducibility and Interoperability}

Another challenge lies in the reproducibility and interoperability of automated workflows. Although the HIKMA pipeline maintains internal traceability through tracking workbooks and artifact hashing, external reproducibility remains limited by differences in model configurations, dataset licensing, and system dependencies. Creating cross-platform reproducibility protocols that allow independent verification of AI-generated manuscripts, reviews, and revisions remains a significant research direction.

\subsection{Evaluation Metrics}

Evaluation metrics also require further refinement. Current quality and safety measures—readability, factuality, citation integrity, and leakage rates—are effective for diagnostic assessment but insufficient for measuring interpretive quality or scholarly value. Future frameworks must include longitudinal measures of research influence, such as citation uptake, community validation, and downstream reuse, to assess whether AI-authored scholarship contributes meaningfully to scientific discourse.

\subsubsection{Expanding Human Input in Future Iterations}

Future iterations of the HIKMA framework will include more deliberate human participation at multiple points in the generation pipeline. Specifically, human experts will be incorporated into the dataset screening process to validate ethical compliance, contextual relevance, and intellectual property alignment before data ingestion. Additionally, subject-area chairs or domain editors will be assigned to review AI-generated manuscripts before they progress to camera-ready status. Their responsibilities will include validating methodological plausibility, interpreting complex findings, and approving final content for archival release. This hybrid structure will maintain efficiency while introducing reflective judgment and disciplinary context that AI systems alone cannot provide. The goal is to evolve from a fully autonomous model to a human-in-the-loop framework where automation and human expertise coexist, ensuring that the scholarly standards of originality, accountability, and interpretive accuracy are preserved.

\subsection{Cultural and Linguistic Inclusivity}

Cultural and linguistic inclusivity present additional challenges. Despite balanced prompting, the outputs reflected patterns of bias linked to model training distributions and linguistic dominance. Ensuring global inclusivity in AI-generated scholarship will require multilingual datasets, regional adaptation of models, and review processes sensitive to cultural context. Without deliberate design for diversity, automation risks reinforcing epistemic asymmetries rather than reducing them.

\subsection{Integration with Human Scholarship}

Integration with human scholarly practice remains a complex issue. The HIKMA experiment shows that AI systems can replicate many procedural aspects of research, but not the reflective judgment or ethical reasoning inherent to human scholarship. Determining how human and machine roles should interact in authorship, review, and interpretation will require long-term study and institutional consensus. Hybrid models, where AI systems handle generation and analysis while human experts validate and contextualize, appear to be the most practical pathway forward.

\subsection{Infrastructure and Sustainability}

Infrastructure and sustainability are emerging concerns. Running large-scale automated scholarly pipelines demands computing resources, data governance systems, and archival infrastructure capable of maintaining audit trails over time. Future work must explore lightweight, distributed, and open-source architectures to make autonomous scholarly frameworks sustainable and accessible beyond large institutions.

\subsection{Social Acceptance and Trust}

Finally, social acceptance and trust remain critical future challenges. Even with transparency and audit mechanisms in place, the legitimacy of AI-authored and AI-reviewed scholarship depends on community perception. Acceptance will require clear communication of how such systems operate, evidence of fairness and rigor, and engagement with researchers, publishers, and policymakers to define appropriate boundaries for automation.

In summary, the HIKMA experiment demonstrated feasibility but also revealed a new research agenda for autonomous scholarly environment. Future work must address governance, reproducibility, inclusivity, human–AI collaboration, infrastructure sustainability, and social legitimacy. These challenges define the next phase of development for transparent, accountable, and trustworthy AI-driven research frameworks.

\section{Credit Author Roles}
\label{sec:credit}
\vspace{0.2cm}

The HIKMA experiment was realized through contributions mapped to the CRediT taxonomy. Conceptualization involved defining the workflow from dataset intake to archival, while methodology established procedures for license checks, manuscript prompting, rubric-based review, revision cycles, and presentation synthesis. Software contributions covered tools for dataset parsing, citation validation, hallucination detection, LaTeX normalization, and artifact hashing, with validation ensuring structural conformity and reproducibility across stages. Formal analysis assessed manuscripts, reviews, and acceptance outcomes, while investigation consisted of dataset selection, manuscript generation, and review execution. Resources included datasets, bibliographic records, computing infrastructure, and platforms for TTS and avatar rendering, and data curation was managed through the multi-sheet tracking workbook. Writing was divided between original drafts produced by \textit{AI Scholar Frontier} and revisions incorporating reviewer feedback, with visualization encompassing slides, narration scripts, and avatar presentations. Supervision ensured compliance with guardrails and workflow order, and project administration coordinated scheduling, track integration, proceedings, and podcast release. Together these roles supported the progression of 30 complete entries through the full scholarly pipeline and their publication on the HIKMA website and podcast platform.

\section{Conclusion}
\vspace{0.2cm}

This paper has outlined the design and implementation of the HIKMA Conference experiment as a complete AI-assisted scholarly pipeline. The process included dataset preparation, manuscript generation, peer review, revision, and presentation, with AI systems supporting each stage under human oversight and intellectual property safeguards.

We offer a reference architecture and governance model for human-supervised semi-autonomous, IP-preserving scholarly communication, validated at HIKMA 2025. The framework is presented for adaptation, benchmarking, and independent audits. These contributions provide a foundation for future work on the integration of AI into research workflows and its implications for scholarly communication.

\section{Acknowledgments}
\vspace{0.2cm}

We acknowledge the use of large language and multimodal models to assist with drafting, analysis, and editing. All outputs were verified and integrated by the authors to ensure accuracy and integrity.

\section{Prototype Demo and Proceedings} 
\vspace{0.2cm}

The complete conference program, including recorded presentations, opening ceremony and panel discussions, is available as a playlist on YouTube: \href{https://www.youtube.com/playlist?list=PLkbLB2E3H350-OaPEyqo1HF3lHqjpOcfS}{HIKMA Conference 2025} \cite{hikma2025_conference_video}. The full proceedings, which include the thirty accepted papers have been archived and are publicly available via Zenodo (DOI: \href{https://doi.org/10.5281/zenodo.17390176}{10.5281/zenodo.17390176}) \cite{hamad_bin_khalifa_university_2025_17390176}.

\bibliographystyle{IEEEtran}
\bibliography{refs}

\begin{thebibliography}{10}
\providecommand{\url}[1]{#1}
\csname url@samestyle\endcsname
\providecommand{\newblock}{\relax}
\providecommand{\bibinfo}[2]{#2}
\providecommand{\BIBentrySTDinterwordspacing}{\spaceskip=0pt\relax}
\providecommand{\BIBentryALTinterwordstretchfactor}{4}
\providecommand{\BIBentryALTinterwordspacing}{\spaceskip=\fontdimen2\font plus
\BIBentryALTinterwordstretchfactor\fontdimen3\font minus \fontdimen4\font\relax}
\providecommand{\BIBforeignlanguage}[2]{{%
\expandafter\ifx\csname l@#1\endcsname\relax
\typeout{** WARNING: IEEEtran.bst: No hyphenation pattern has been}%
\typeout{** loaded for the language `#1'. Using the pattern for}%
\typeout{** the default language instead.}%
\else
\language=\csname l@#1\endcsname
\fi
#2}}
\providecommand{\BIBdecl}{\relax}
\BIBdecl

\bibitem{floridi2020gpt}
L.~Floridi and M.~Chiriatti, ``Gpt-3: Its nature, scope, limits, and consequences,'' \emph{Minds and machines}, vol.~30, no.~4, pp. 681--694, 2020.

\bibitem{gil2022will}
Y.~Gil, ``{Will AI write scientific papers in the future?}'' \emph{AI Magazine}, vol.~42, no.~4, pp. 3--15, 2022.

\bibitem{salvagno2023can}
M.~Salvagno, F.~S. Taccone, and A.~G. Gerli, ``{Can artificial intelligence help for scientific writing?}'' \emph{Critical care}, vol.~27, no.~1, p.~75, 2023.

\bibitem{auernhammer2020human}
J.~Auernhammer, ``{Human-centered AI: The role of Human-centered Design Research in the development of AI},'' 2020.

\bibitem{afzal2025chatgpt}
M.~Afzal, N.~Arshad, and A.~Shaheen, ``{ChatGPT and the Future of Academic Writing: Enhancing Productivity and Creativity},'' \emph{Journal of Engineering and Computational Intelligence Review}, vol.~3, no.~1, pp. 1--11, 2025.

\bibitem{donovan2018algorithmic}
J.~Donovan, R.~Caplan, J.~Matthews, and L.~Hanson, ``{Algorithmic accountability: a primer},'' 2018.

\bibitem{li2023trustworthy}
B.~Li, P.~Qi, B.~Liu, S.~Di, J.~Liu, J.~Pei, J.~Yi, and B.~Zhou, ``{Trustworthy AI: From principles to practices},'' \emph{ACM Computing Surveys}, vol.~55, no.~9, pp. 1--46, 2023.

\bibitem{weidinger2021ethical}
L.~Weidinger, J.~Mellor, M.~Rauh, C.~Griffin, J.~Uesato, P.-S. Huang, M.~Cheng, M.~Glaese, B.~Balle, A.~Kasirzadeh \emph{et~al.}, ``{Ethical and social risks of harm from language models},'' \emph{arXiv preprint arXiv:2112.04359}, 2021.

\bibitem{yoo2024evolving}
Y.~Yoo, ``{Evolving epistemic infrastructure: The role of scientific journals in the age of generative AI},'' \emph{Journal of the Association for Information Systems}, vol.~25, no.~1, pp. 137--144, 2024.

\bibitem{heygen2022}
{HeyGen Technology, Inc.}, ``Heygen: Ai-powered video generation and avatar creation platform,'' \url{https://www.heygen.com}, Los Angeles, CA, 2022, aI video platform with voice cloning and avatar generation capabilities. Accessed: 2025-10-04.

\bibitem{swanson1986undiscovered}
D.~R. Swanson, ``{Undiscovered public knowledge},'' \emph{The Library Quarterly}, vol.~56, no.~2, pp. 103--118, 1986.

\bibitem{sparkes2010towards}
A.~Sparkes, R.~D. King \emph{et~al.}, ``{Towards robot scientists for autonomous scientific discovery},'' \emph{Automated Experimentation}, vol.~2, no.~1, pp. 1--9, 2010.

\bibitem{mccall2025ai}
A.~McCall and A.~Mccall, ``{AI for Scientific Discovery: Automating Hypothesis Generation},'' 2025.

\bibitem{nathani2025can}
K.~R. Nathani, A.-M. Nathani, M.~Delawan, A.~Safdar, and M.~Bydon, ``{Can artificial intelligence write science? A comparative analysis of human-written and artificial intelligence--generated scientific writings},'' \emph{Journal of Neurosurgery: Spine}, vol.~1, no. aop, pp. 1--6, 2025.

\bibitem{elicit2023}
``{Elicit: The AI research assistant},'' 2023, available at \url{https://elicit.org}.

\bibitem{whitfield2023elicit}
S.~Whitfield and M.~A. Hofmann, ``{Elicit: AI literature review research assistant},'' \emph{Public Services Quarterly}, vol.~19, no.~3, pp. 201--207, 2023.

\bibitem{writefull2022}
``{Writefull: AI writing support for researchers},'' 2022, available at \url{https://writefull.com}.

\bibitem{bute2025writefull}
A.~R. Bute and D.~K. Ambast, ``{Writefull’s contribution to advancing writing capabilities: A review},'' in \emph{AIP Conference Proceedings}, vol. 3298, no.~1.\hskip 1em plus 0.5em minus 0.4em\relax AIP Publishing LLC, 2025, p. 040026.

\bibitem{scholarcy2022}
``{Scholarcy: The AI-powered article summarizer},'' 2022, available at \url{https://scholarcy.com}.

\bibitem{bui2024decoding}
T.~X.~H. Bui and V.~H. Bui, ``{Decoding Scholarcy website: A Study on its Research Summarization Efficiency},'' in \emph{Proceedings of the AsiaCALL International Conference}, vol.~6, 2024, pp. 71--80.

\bibitem{co_scientist2024}
``{Google DeepMind Co-Scientist: AI for automated hypothesis testing and discovery},'' 2024, available at \url{https://deepmind.google/discover/blog/introducing-co-scientist}.

\bibitem{sakana2023}
``{Sakana AI Scientist: Autonomous research agent},'' 2023, available at \url{https://sakana.ai}.

\bibitem{stanford2025agents4science}
{Stanford University}, ``Agents for science: An open conference on ai-generated research and peer review,'' \url{https://agents4science.stanford.edu/}, Stanford University, Department of Computer Science and Center for Research on Foundation Models (CRFM), 2025, accessed: 2025-10-04.

\bibitem{zhang2025aixiv}
P.~Zhang, X.~Hu, G.~Huang, Y.~Qi, H.~Zhang, X.~Li, J.~Song, J.~Luo, Y.~Li, S.~Yin \emph{et~al.}, ``{A Next-Generation Open Access Ecosystem for Scientific Discovery Generated by AI Scientists},'' \emph{arXiv preprint arXiv:2508.15126}, 2025.

\bibitem{checco2021ai}
A.~Checco, L.~Bracciale, P.~Loreti, S.~Pinfield, and G.~Bianchi, ``{AI-assisted peer review},'' \emph{Humanities and social sciences communications}, vol.~8, no.~1, pp. 1--11, 2021.

\bibitem{donker2023dangers}
T.~Donker, ``{The dangers of using large language models for peer review},'' \emph{The Lancet Infectious Diseases}, vol.~23, no.~7, p. 781, 2023.

\bibitem{mitchell2021algorithmic}
S.~Mitchell, E.~Potash, S.~Barocas, A.~D'Amour, and K.~Lum, ``{Algorithmic fairness: Choices, assumptions, and definitions},'' \emph{Annual review of statistics and its application}, vol.~8, no.~1, pp. 141--163, 2021.

\bibitem{mitchell2019model}
M.~Mitchell \emph{et~al.}, ``{Model cards for model reporting},'' \emph{Proceedings of the Conference on Fairness, Accountability, and Transparency}, pp. 220--229, 2019.

\bibitem{gebru2021datasheets}
T.~Gebru \emph{et~al.}, ``{Datasheets for datasets},'' \emph{Communications of the ACM}, vol.~64, no.~12, pp. 86--92, 2021.

\bibitem{stodden2018enhancing}
V.~Stodden \emph{et~al.}, ``{Enhancing reproducibility for computational methods},'' \emph{Science}, vol. 354, no. 6317, pp. 1240--1241, 2018.

\bibitem{peng2011reproducible}
R.~D. Peng, ``{Reproducible research in computational science},'' \emph{Science}, vol. 334, no. 6060, pp. 1226--1227, 2011.

\bibitem{shen2018natural}
J.~Shen \emph{et~al.}, ``{Natural TTS synthesis by conditioning WaveNet on Mel spectrogram predictions},'' in \emph{ICASSP}, 2018, pp. 4779--4783.

\bibitem{van2016wavenet}
A.~van~den Oord \emph{et~al.}, ``{WaveNet: A generative model for raw audio},'' \emph{arXiv preprint arXiv:1609.03499}, 2016.

\bibitem{kim2021conditional}
J.~Kim \emph{et~al.}, ``{Conditional variational autoencoder with adversarial learning for end-to-end text-to-speech},'' \emph{arXiv preprint arXiv:2106.06103}, 2021.

\bibitem{hikma2025_opening_remarks}
{HIKMA Project}, ``Hikma conference opening remarks,'' \url{https://www.youtube.com/watch?v=qD2Hb65f69s&list=PLkbLB2E3H350-OaPEyqo1HF3lHqjpOcfS&index=3}, Hamad Bin Khalifa University (HBKU), College of Science and Engineering, 2025.

\bibitem{hikma2025_avatar_video}
------, ``Hikma 2025 | virtual avatar talk: From minimal state to welfare state (progressive education track),'' \url{https://www.youtube.com/watch?v=s5OkcTiKmcw&list=PLkbLB2E3H350-OaPEyqo1HF3lHqjpOcfS&index=4}, Hamad Bin Khalifa University (HBKU), College of Science and Engineering, 2025.

\bibitem{hikma2025_podcast}
------, ``Hikma podcast: The future of ai in academia,'' \url{https://www.youtube.com/watch?v=s5OkcTiKmcw&list=PLkbLB2E3H350-OaPEyqo1HF3lHqjpOcfS&index=4}, Hamad Bin Khalifa University (HBKU), College of Science and Engineering, 2025, accessed: 2025-10-04.

\bibitem{google_notebooklm}
{Google DeepMind}, ``{NotebookLM: AI-Powered Note and Research Tool},'' \url{https://notebooklm.google.com/}, Google LLC, 2024, accessed: 2025-10-04.

\bibitem{canva_tool}
{Canva Pty Ltd.}, ``{Canva: Online Graphic Design Platform},'' \url{https://www.canva.com/}, Canva Pty Ltd., 2025, accessed: 2025-10-04.

\bibitem{hikma2025_conference_video}
{HIKMA Project}, ``Hikma conference 2025 walkthrough,'' \url{https://www.youtube.com/watch?v=eszMLHw7Ijo&list=PLkbLB2E3H350-OaPEyqo1HF3lHqjpOcfS&index=4}, Hamad Bin Khalifa University (HBKU), College of Science and Engineering, 2025, youTube, accessed: 2025-10-04.

\bibitem{hamad_bin_khalifa_university_2025_17390176}
\BIBentryALTinterwordspacing
{Hamad Bin Khalifa University}, ``Hikma conference 2025 — proceedings: Thirty accepted papers, slide decks, recorded presentations, and podcast series,'' Oct. 2025. [Online]. Available: \url{https://doi.org/10.5281/zenodo.17390176}
\BIBentrySTDinterwordspacing

\end{thebibliography}

\appendices

\clearpage
\onecolumn

\section{Paper Generation Tracking Worksheet} 
\label{app:worksheet}



\definecolor{Gray1}{gray}{0.8}
\definecolor{Gray2}{gray}{0.9}

\begin{table*}[h!]
\centering
\small
\caption{HIKMA Pipeline Worksheet: Dataset, Generation Metrics, Review Process, and Presentation Status (Social Progress Track)}
\label{tab:hikma_worksheet}

\renewcommand{\arraystretch}{1.1}

\rotatebox{90}{
    \resizebox{0.85\textheight}{!}{  
    \begin{tabular}{|p{3.3em}|p{3.5em}|p{13em}|p{7em}
                    |p{2.5em}|p{2em}|p{3em}|p{3em}
                    |p{3em}|p{3em}|p{2.7em}|p{3.5em}|p{4em}
                    |p{3.5em}|}
    \hline
    \rowcolor{Gray1}
    \textbf{Track} & \textbf{Paper ID} & \textbf{Dataset (link)} &
    \textbf{Output Paper Title} &
    \textbf{Pages} & \textbf{Time (hrs)} &
    \textbf{Rev 1 Score} & \textbf{Rev 1 Decision} &
    \textbf{Rev 2 Score} & \textbf{Rev 2 Decision} &
    \textbf{Total Score (0--10)} &
    \textbf{Initial Decision} &
    \textbf{Revised Paper / Response Letter} &
    \textbf{Final Decision} \\
    \hline

    Social Progress & PT1-SOCP-01 &
    \url{https://www.kaggle.com/datasets/nandinivishwanathan/social-progress-index/} &
    \textbf{Beyond GDP: A Multi-Dimensional Analysis of Global Social Progress} &
    10 & 1.5 & 7.0 & Weak Accept & 8.0 & Weak Accept & 7.5 & \textsc{Accept} & YES / YES & ACCEPT \\
    \hline

    Social Progress & PT1-SOCP-02 &
    \url{https://www.kaggle.com/datasets/itsujitsharma/student-stress-monitoring-dataset} &
    \textbf{Beyond the Breaking Point: Mapping Stress Dynamics \& Resilience Pathways in Students} &
    9 & 1.5 & 8.0 & Weak Accept & 8.0 & Weak Accept & 8.0 & \textsc{Accept} & YES / YES & ACCEPT \\
    \hline

    Social Progress & PT1-SOCP-03 &
    \url{https://stats.oecd.org/Index.aspx?DataSetCode=BLI} &
    \textbf{Analysis of Well-Being Across OECD Countries Using the Better Life Index} &
    7 & 1.5 & 7.0 & Accept & 4.0 & \textbf{Reject} & 5.5 & \textsc{Reject} & NO / NO & REJECT \\
    \hline

    Social Progress & PT1-SOCP-04 &
    \url{https://www.kaggle.com/datasets/unsdsn/world-happiness} &
    \textbf{Beyond GDP: Social Connections \& Freedom as Drivers of Wellbeing (WHR 2024)} &
    9 & 1.5 & 7.0 & Weak Accept & 5.0 & \textbf{Reject} & 6.0 & \textsc{Reject} & NO / NO & REJECT \\
    \hline

    Social Progress & PT1-SOCP-05 &
    \url{https://www.kaggle.com/datasets/l3llff/japan-life-expectancy} &
    \textbf{Upstream Drivers of Longevity: Education, Earnings, \& Work-Time in Japan} &
    8 & 1.5 & 8.0 & Accept & 5.0 & \textbf{Reject} & 6.5 & \textsc{Reject} & NO / NO & REJECT \\
    \hline
    \end{tabular}
    } 
} 
\end{table*}




        
        
        
        
        
        
        
        
        
        

      



\clearpage
\section{Prompt Library}  

\vspace{0.3cm}

\subsection{Input Data Prompt for Paper Generation through AI Scholar Frontier} 
\begin{figure} [h]
  \centering
  \includegraphics[width=0.9\linewidth]{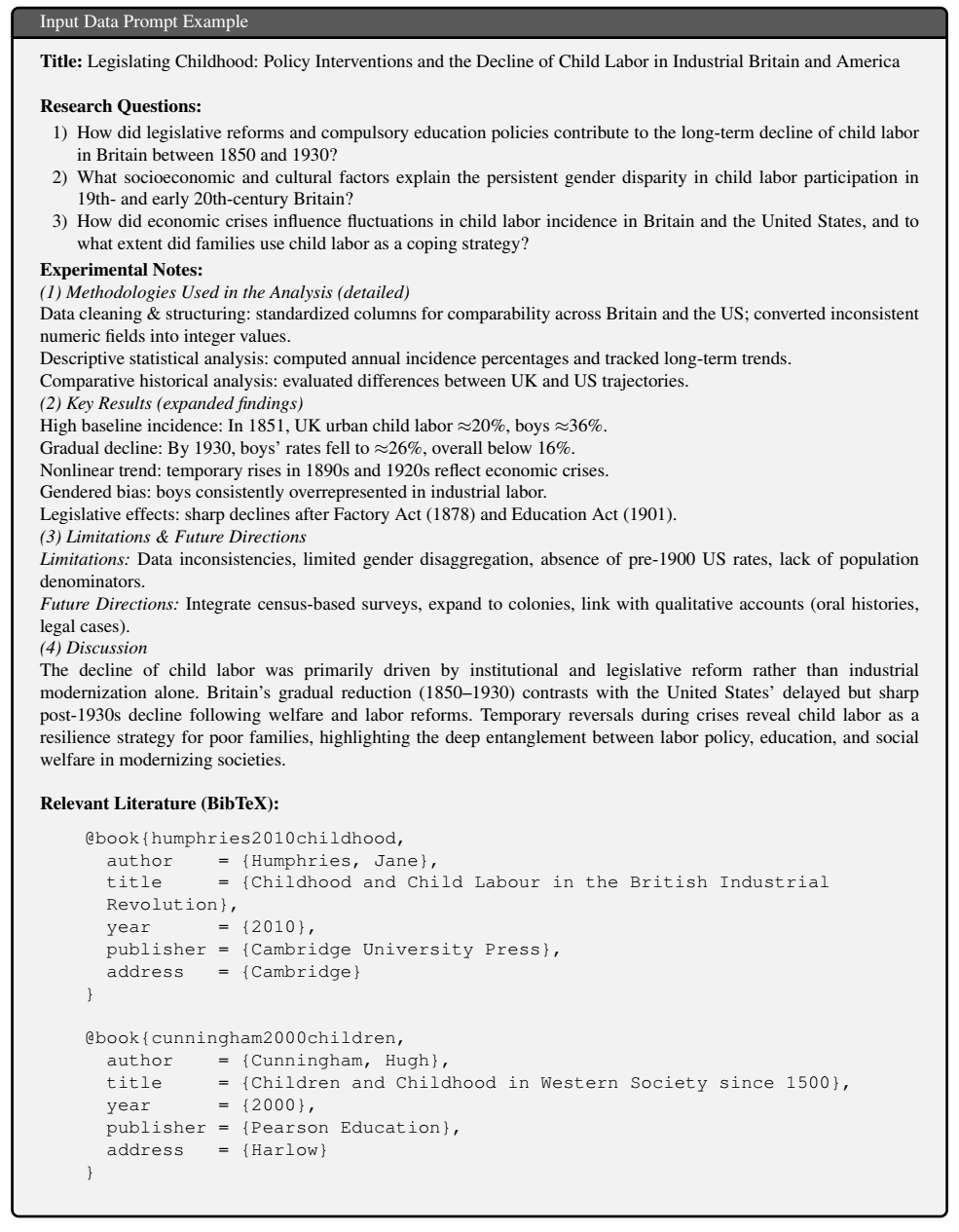}
  \label{app:input_prompt}
\end{figure} 

\clearpage
\subsection{First Reviewer Prompt}
\begin{figure} [h]
  \centering
  \includegraphics[width=0.9\linewidth]{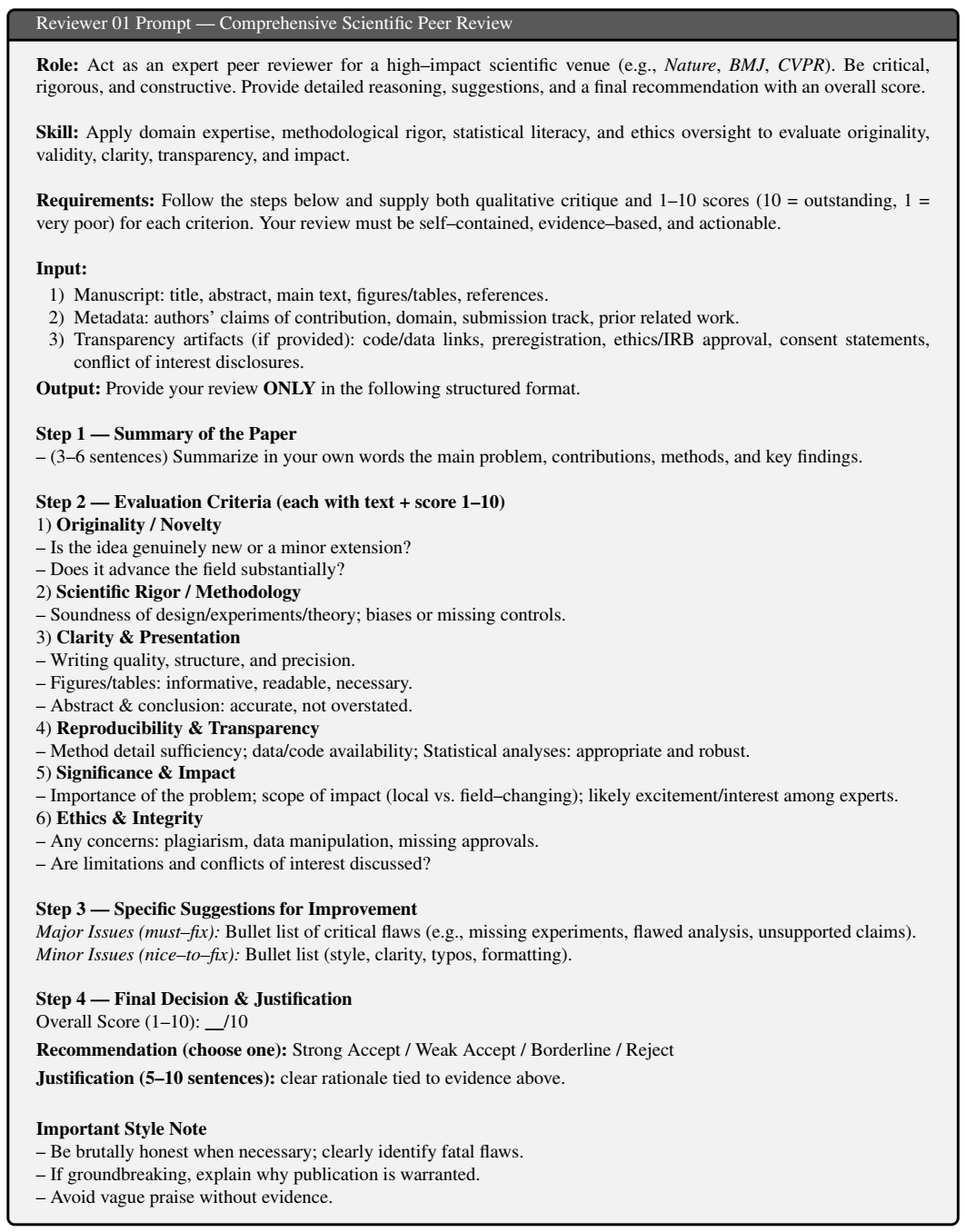}
  \label{app:reviewer_prompt_01} 
\end{figure} 

\clearpage
\subsection{Second Reviewer Prompt}
\begin{figure} [h]
  \centering
  \includegraphics[width=0.9\linewidth]{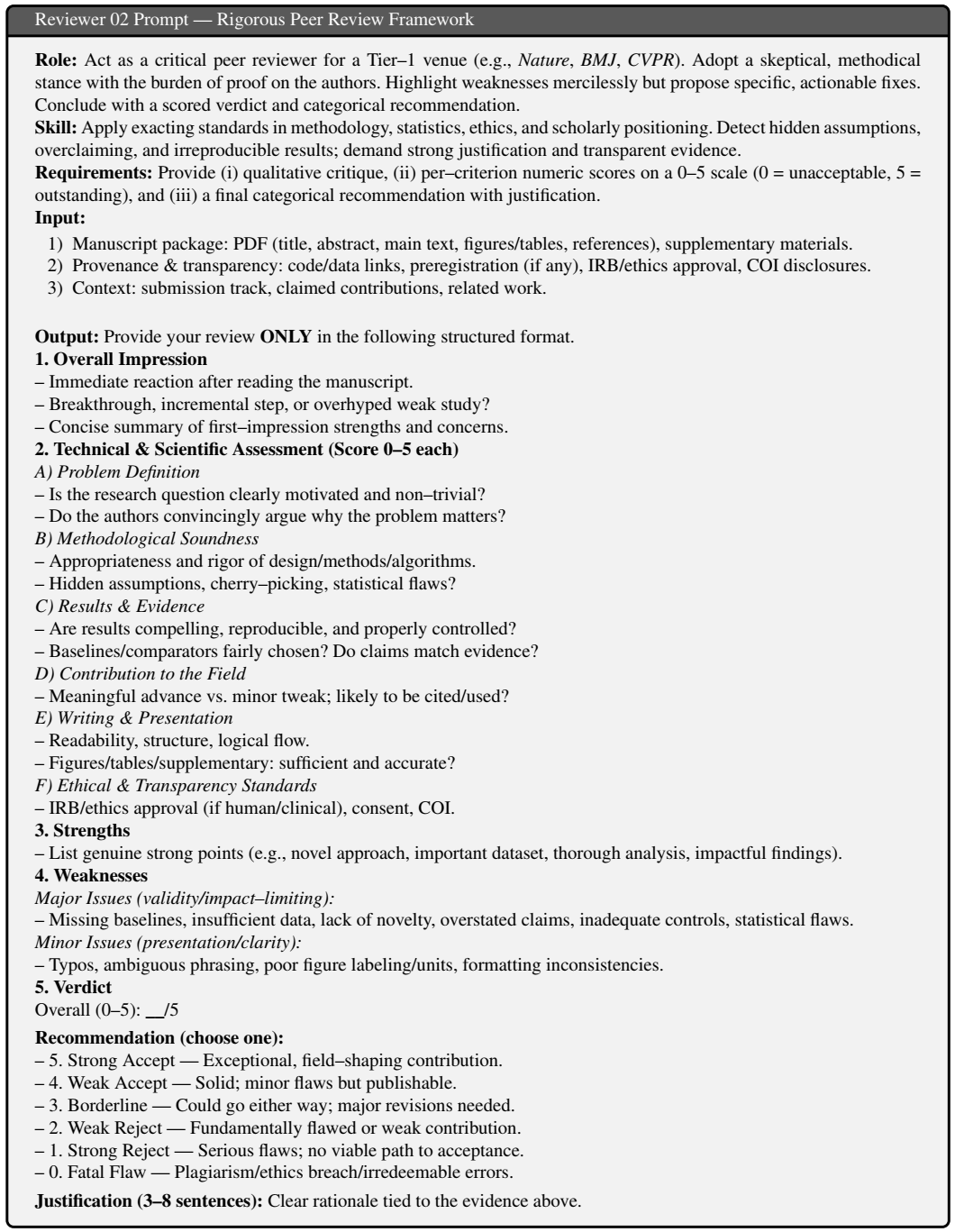}
  \label{app:reviewer_prompt_02} 
\end{figure} 

\clearpage
\subsection{Revision Task Prompt}

\begin{figure} [h]
  \centering
  \includegraphics[width=0.9\linewidth]{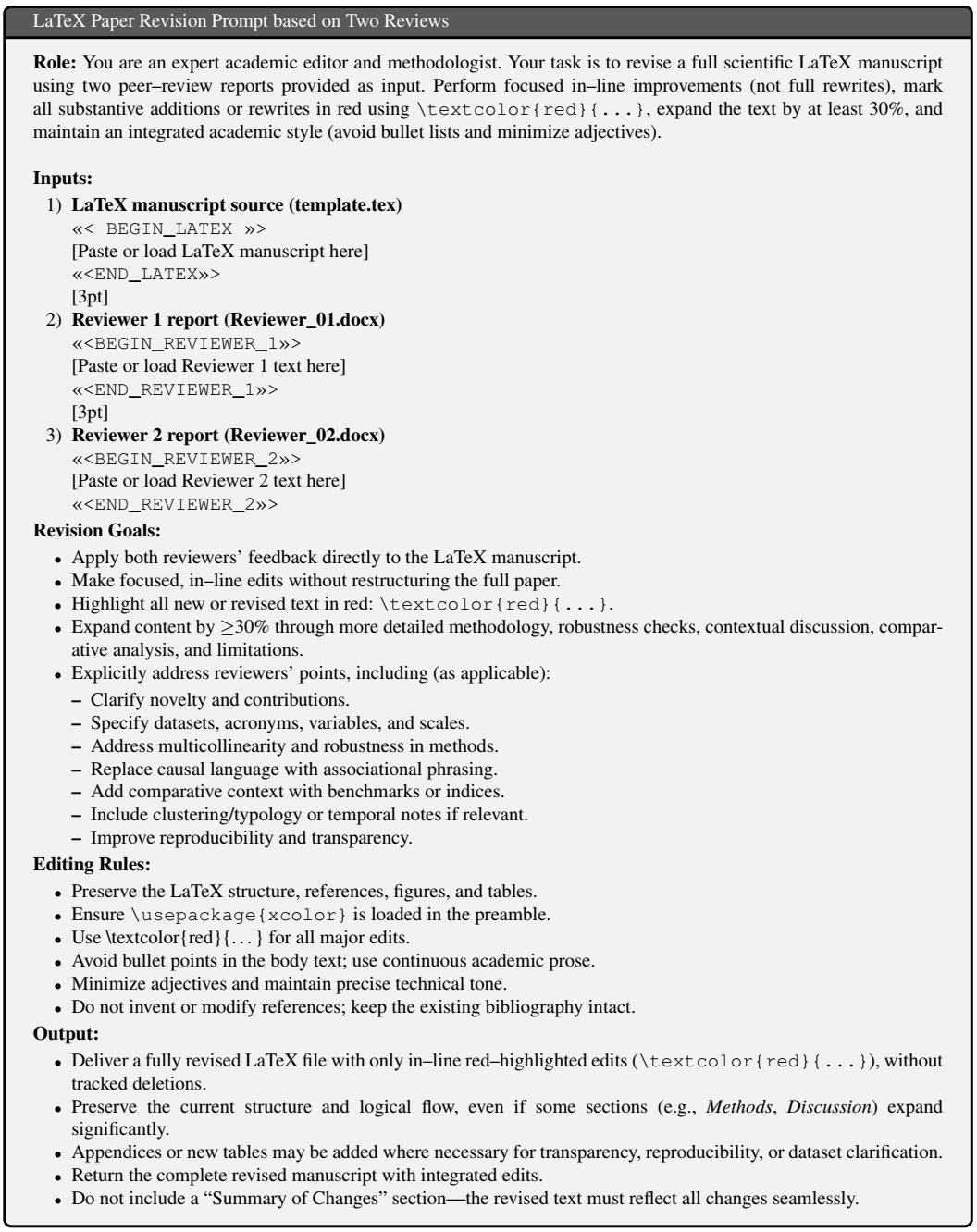}
  \label{app:revision_prompt} 
\end{figure} 

\clearpage

\subsection{Response or Rebuttal Letter Prompt}

\begin{figure} [h]
  \centering
  \includegraphics[width=0.88\linewidth]{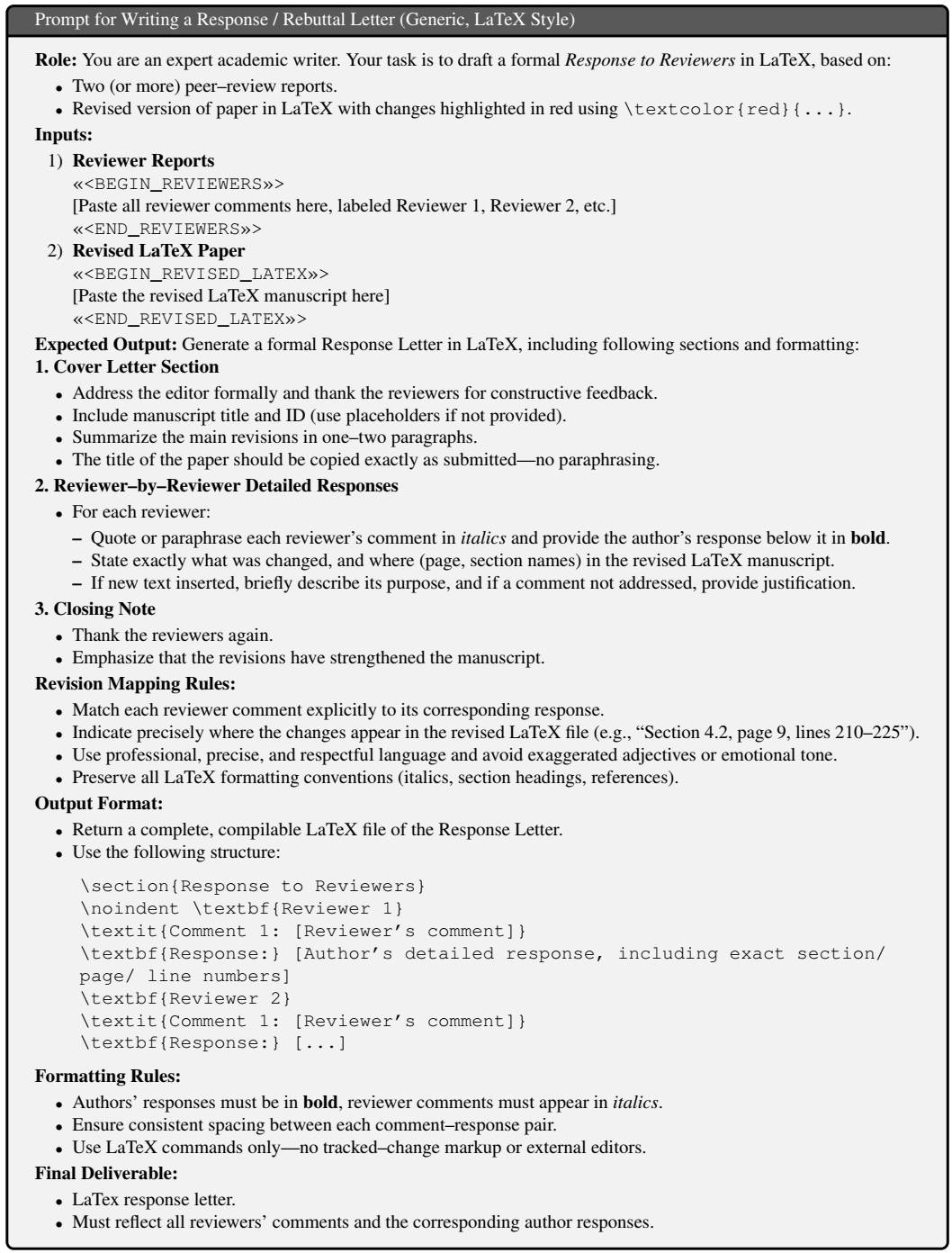}
  \label{app:response_letter_promnpt}
\end{figure} 

\clearpage
\twocolumn

\begin{IEEEbiography}{ZAIN  TARIQ}, (Member, IEEE) received his M.S. degree in Computer and Communications Security from the National University of Sciences and Technology (NUST) in 2014 with a Gold Medal and his Ph.D. in Computer Science and Engineering (CSE) from the College of Science and Engineering (CSE), Hamad Bin Khalifa University (HBKU), Doha – Qatar in 2024. He currently holds a Post-Doctoral Researcher position at Hamad Bin Khalifa University. His research interests include affective computing, cybersecurity, wireless communications, large language models and reinforcement learning. Contact him at ztariq@hbku.edu.qa.
\vspace*{8pt}
\end{IEEEbiography}

\begin{IEEEbiography}{Mahmood ALzubaidi} completed his Master's degree in 2018 in Internet Engineering from the National Advanced IPv6 Center at Universiti Sains Malaysia. He then pursued and received his PhD in 2023 from Hamad Bin Khalifa University, Qatar, where he continues to contribute to the field as a Post-Doc Researcher. His research interests are broad, spanning across the Internet of Things (IoT), deep learning, machine learning, image segmentation, generative AI, and AI application in healthcare. Contact him as malzubaidi@hbku.edu.qa
\vspace*{8pt}
\end{IEEEbiography}

\begin{IEEEbiography}{Uzair Shah} is a Ph.D. candidate in Computer Science and Engineering at the College of Science and Engineering, Hamad Bin Khalifa University (HBKU), Doha, Qatar. His research focuses on artificial intelligence for medical imaging, multimodal clinical data analysis, and computer vision. He completed his Master’s degree in Data Analytics from Hamad Bin Khalifa University, Doha, Qatar. His current work explores generative AI for brain MRI augmentation, multimodal fusion of electronic health records and imaging data, and medical image segmentation. He has published in CVPR Workshops and leading biomedical journals and serves on the editorial team for MedInfo 2025 proceedings. His professional experience includes research roles at KAUST (VSRP Program) and PanData, Qatar, where he worked on predictive healthcare modeling. His research interests include 3D segmentation, visualization, and vision–language models. 
Contact him at uzsh31989@hbku.edu.qa.
\vadjust{\vfill\pagebreak}
\end{IEEEbiography}

\begin{IEEEbiography}{Marco Agus} is an Associate Professor in the College of Science and Engineering at Hamad Bin Khalifa University. He obtained an MSc and PhD from the University of Cagliari, Italy. He has also worked as a Research Engineer at the King Abdullah University of Science and Technology, Saudi Arabia, and as a Research Scientist at the Center of Research, Development, and Advanced Studies (CRS4), in Cagliari, Italy.
Marco Agus's research interests span different domains in visual computing, from haptics and visual rendering for medical applications to real-time exploration of massive models, to machine learning methods for electron microscopy biology data and indoor environments. He taught courses at several important visual computing venues, including IEEE CVPR, ACM SIGGRAPH, and Eurographics, and he regularly acts as a committee member, reviewer, chair, and associate editor for top journals and conferences in the visual computing domain. Contact him at magus@hbku.edu.qa.
\vspace*{7pt}
\end{IEEEbiography}

\begin{IEEEbiography}{Mowafa Said Househ (PhD)} is a Full Professor of Health Informatics in the Arab World and is associated with the College of Science and Engineering, Hamad Bin Khalifa University, Doha, Qatar. His research interests include the use of information and communication technologies in healthcare, with a particular focus on Artificial Intelligence Technologies to support public health and improve healthcare literacy. He has a strong interest in the field of digital mental health, aiming to use technology to benefit both patients and healthcare professionals. In 2020, Prof. Househ was the recipient of first place in the United Nations ITU Artificial Intelligence Challenge. His work has also been recognized with the Researcher Award from the King Abdullah International Medical Research Centre in 2015, the MBC Hope Awards for Innovation and Social Entrepreneurship in 2015, and as a finalist for the Massachusetts Institute of Technology Arab Enterprise Forum competition in 2014.Additional recognitions include the Young Researcher Award in 2010, the Northern British Columbia Healthcare in Technology Award in 2007, and the Michael Smith Foundation for Health Research Trainee Award in 2003.Prof. Househ also holds the position of adjunct Professor at the University of Victoria's School of Health Information Science in Victoria, BC, Canada. With a scholarly contribution of over 340 research papers and more than 14 books, Prof. Househ's work in health informatics has been influential in both the academic realm and practical application, impacting healthcare practices in the Arab world and internationally. Contact him at mhouseh@hbku.edu.qa.
\vspace*{8pt}
\end{IEEEbiography}

\end{document}